\documentclass[prd,aps,preprint,amsmath,nofootinbib,amssymb,eqsecnum,showkeys,tightenlines]{revtex4-1}
\usepackage{tabularx}
\pdfoutput=1
\usepackage{slashed}
\usepackage{epsfig,latexsym,cancel,amssymb,amsmath,verbatim,mathrsfs}
\usepackage{color}
\usepackage{graphicx}
\usepackage{bm}
\usepackage{tabularx}
\usepackage{amsmath}
\usepackage{amssymb}
\usepackage{amsfonts}
\usepackage{cases}
\usepackage{cancel}
\usepackage{hyperref}
\usepackage{ulem}
\usepackage{here}
\usepackage{color}
\usepackage{graphicx}
\usepackage{diagbox}
\usepackage{multirow}
\usepackage{slashed}
\usepackage{simpler-wick}
\usepackage{subfigure}
\usepackage{svg}
\usepackage[marginal]{footmisc}
\usepackage{feynmf}
\usepackage{array}
\usepackage[T1]{fontenc}
\usepackage{multirow}
\usepackage{booktabs}
\usepackage{xspace}
\usepackage{mathrsfs}
\newcommand{\vt}{\texttt{VacuumTunneling}\@\xspace}
\newcommand{\mth}{\texttt{Mathematica}\@\xspace}
\newcommand{\tul}{\texttt{Tunneling}\@\xspace}
\definecolor{mio}{RGB}{59,110,147}
\definecolor{mvar}{RGB}{67,137,88}
\definecolor{mpar}{RGB}{60,125,145}

\def\d{\mathrm{d}}

\def\tanh{\rm Tanh}

\def\wt{\widetilde}

\newcommand{\be}{\begin{equation}}
	\newcommand{\ee}{\end{equation}}
\newcommand{\bea}{\begin{eqnarray}}
	\newcommand{\eea}{\end{eqnarray}}
\newcommand{\ba}{\begin{array}}
	\newcommand{\ea}{\end{array}}

\def\wt{\widetilde}

\long\def\symbolfootnote[#1]#2{\begingroup%
\def\thefootnote{\fnsymbol{footnote}}\footnote[#1]{#2}\endgroup}

\newcommand{\beq}{\begin{equation}}
	\newcommand{\eeq}{\end{equation}}

\begin{document}

\title{VacuumTunneling: A package to solve bounce equation with renormalization factor}

\author{Banghui Hua}
\email{bhhua@hust.edu.cn}
\affiliation{School of Physics, Huazhong University of Science and Technology, Wuhan 430074, China}

\author{Jiang Zhu}
\email{jackpotzhujiang@gmail.com}
\affiliation{Tsung-Dao Lee Institute and  School of Physics and Astronomy, Shanghai Jiao Tong University,
	800 Dongchuan Road, Shanghai, 200240 China}
\affiliation{Shanghai Key Laboratory for Particle Physics and Cosmology, 
	Key Laboratory for Particle Astrophysics and Cosmology (MOE), 
	Shanghai Jiao Tong University, Shanghai 200240, China}

\date{\today}

\begin{abstract}

The Vacuum tunneling rate $\Gamma$ from the effective action is a key to studying the cosmological first-order phase transition(FOPT). One solid way to compute the $\Gamma$ is to start with the derivative expansion of the effective action and solve the bounce equation numerically. In this process, the renormalization factor $Z$ of the tunneling field may play a center rule, which is not considered in existing packages.
Therefore, we present a \texttt{Mathematica} package \vt to compute the bounce action 
with or without the renormalization factor. Applying the \vt package, 
we find that the presence of $Z$ has a significant impact on the action, 
as well on the tunneling path. We provide some concrete examples to demonstrate the difference between the solution with and without the renormalization factor, both in the action and tunneling path.
This package is based on the modified shooting and path deformation method. 
We also made some optimizations for the super-cooling phase transition(thick wall scenario), in which other numerical package works poorly. This package works as long as the expressions can give values of the potential and the renormalization at a certain field point. This means the input potential and the renormalization can be merely numerical quantities without analytical expressions. The computation time can be as short as 1 second in single-field tunneling and several seconds in multi-field cases.

\end{abstract}

\pacs{12.60.Jv,  14.70.Pw,  95.35.+d}

\maketitle

\clearpage
\section*{Program Summary}
\noindent
{\em Program title:} \texttt{VacuumTunneling} \\[0.5em]
{\em Program obtainable from:} \url{https://github.com/bhhua/VacuumTunneling} \\[0.5em]
{\em Distribution format:} .wl \\[0.5em]
{\em Programming language:} \texttt{Mathematica} \\[0.5em]
{\em Computer:} Personal computer \\[0.5em]
{\em Operating system:} Tested on Windows, should work wherever \texttt{Mathematica} is installed. \\[0.5em]
{\em Typical running time:} O(1) second for 1 field. O(10) seconds for multi fields. \\[0.5em]
{\em Nature of problem:}  Evaluation of the Euclidean bounce action that controls the decay rate including a wave-function renormalization.\\[0.5em]
{\em Solution method:}  Shooting method and path deformation.\\[0.5em]
{\em Restrictions:} \texttt{Mathematica} version 12.2 or above, works in $D=3,4$. \\[0.5em]

\clearpage
\tableofcontents

\newpage
\section{Introduction}\label{Intro}


The structure of particle physics is driven by the symmetries, which is especially important to build beyond standard models~\cite{Sojka:2024btp,Hambye:2013dgv,Jinno:2016knw,Marzola:2017jzl,Hashino:2018zsi,Prokopec:2018tnq} physics. Although it is essential, it may also bring some changes when considering the history of our universe. Those underlying symmetries may be preserved when the universe's temperature is quite large but spontaneously broken at zero temperature theory~\cite{Dolan:1973qd, Weinberg:1974hy, Sher:1988mj, Kang:2020jeg,Ellis:2020nnr,Konstandin:2010cd,Azatov:2020nbe}. This process would cause a phase transition and change the equation of the state of the universe. If the phase transition is a first-order phase transition(FOPT), it may also produce the stochastic gravitational background signal(GWs)~\cite{Witten:1984rs,Hogan:1986dsh,Kosowsky:1991ua,Kosowsky:1992rz,Grojean:2006bp,Hindmarsh:2013xza,Hindmarsh:2015qta,Hindmarsh:2017gnf,Cutting:2018tjt,Cutting:2019zws,Wang:2020jrd,Athron:2023xlk} and can be observed in further space-based GWs detector~\cite{Crowder:2005nr,Corbin:2005ny,Harry:2006fi,Caprini:2015zlo,LISA:2017pwj,Caprini:2019egz,TianQin:2015yph,TianQin:2020hid,Ruan:2018tsw,Ruan:2020smc}. The phase transition may also help us solve some long-standing mysteries of our universe, such as baryon asymmetry via electroweak baryogenesis~\cite{Cohen:1993nk,Rubakov:1996vz,Riotto:1999yt,Dine:2003ax,Cline:2006ts,Morrissey:2012db,Beniwal:2018hyi,Song:2022xts}. 

On the other hand, describing the FOPT requires understanding the scalar field's quantum tunneling system, with the tunneling field which is the order parameter that labels the vacuum. Coleman and his coworkers first established those general descriptions at zero temperature~\cite{Coleman:1977py, Callan:1977pt, Weinberg:1992ds,Andreassen:2014eha} and called those tunneling rates the false vacuum decay rate $\Gamma$. After that, Linde and others generate that formalism into the finite temperature system~\cite{Linde:1977mm, Linde:1980tt, Linde:1981zj, Quiros:1999jp} or even curved spacetime~\cite{Coleman:1980aw,Hiscock:1987hn,Samuel:1992wt,Cheung:2013sxa,Gregory:2013hja}. Even though this issue has persisted for a long time, numerous recent developments have surfaced in this field, including Green's function-based methods and functional real-time formalism~\cite{Andreassen:2016cff,Andreassen:2016cvx,Ai:2019fri,Ai:2020sru}. The three-dimensional effective field theory is also useful for examining the tunneling rate in both perturbative calculations and lattice simulations~\cite{Gould:2019qek,Croon:2020cgk,Gould:2021oba, Moore:2000jw, Bai:2024pii}. As a gauge invariant quantity, it would affect all observable in the phase transitions. So, it is essential to understand the computation of the tunneling rate $\Gamma$. In general, the tunneling rate $\Gamma$ can be formulated as
\begin{equation}
    \Gamma=A e^{-B},
\end{equation}
where $A$ is the functional determinant of the quantum fluctuation, and $B$ is the so-called bounce action. Obviously, the dominant contribution is from the bounce action $B$, which has an exponential enhancement. The difficulty in calculating $B$ is that one must solve a partial differential equation(PDE) by fixing boundary conditions but having an unknown initial value. This problem is quite complicated in a multi-field tunneling situation. 

There are several numerical methods and packages to compute the bounce action $B$: 
Using shooting method for single field tunneling, there is a package \texttt{CosmoTransitions}~\cite{Wainwright:2011kj} evaluating multi-field tunneling by path deformation and another package \texttt{BubbleProfiler}~\cite{Athron:2019nbd,Athron:2020sbe,Athron:2024xrh} by field perturbative method. The multiple shooting method is used in the package \texttt{AnyBubble}~\cite{Olum:2016bed,Masoumi:2016wot,Masoumi:2017trx}.
Gradient flow method~\cite{Claudson:1983et,Sato:2019axv,Hong:2023dan}, which transforms the bounce solution from a saddle point of the action into a minimum, is used in the package \texttt{SimpleBounce}~\cite{Sato:2019wpo}. The package \texttt{BubbleDet}~\cite{Ekstedt:2023sqc} evaluates the bounce action by decomposing the potential into spherical harmonics. The multi-field tunneling rate can be found by iterating multiple single-field solutions to minimize the action~\cite{Moreno:1998bq,John:1998ip}. There are also methods that treat the bounce as a continuation of an undamped solution~\cite{Park:2010rh,Konstandin:2006nd}. 

Although those numerical packages can compute the bounce action $B$ with quite a high accuracy, a theoretical problem still exists. If the potential in the tunneling problem exists at the classical level, just like the quantum mechanics tunneling problem, then those packages work fine. However, the FOPT is usually induced in the real universe by considering the quantum correction, which means one must consider the tunneling potential at the next leading order(NLO). To make the result theoretically consistent, one has to consider the kinetic terms with the wave function renormalization factor in the theory at the NLO. If the NLO kinetic energy term is not considered, several problems would be induced, such as the gauge dependence of the bounce action~\cite{Metaxas:1995ab,Garny:2012cg,DiLuzio:2014bua,Lofgren:2021ogg,Hirvonen:2021zej}, and make the whole result unreliable. As a requirement of theoretical self-consistency, we must solve real tunneling problems based on effective action. Consequently, the PDE that must be solved differs from the target for which those numerical packages are designed. 

Therefore, it is necessary to consider and design a package to compute $B$ at NLO. For the first time, this paper presents an easy-to-use \texttt{Mathematica} public package based on the shooting-overshooting method for computing the bounce action $B$ at NLO. It can compute both the result with and without the wave function renormalization factors at any dimension. By simply providing the expressions for the effective potential and renormalization and location of the vacua to the main function \texttt{Tunneling}, the program would output the action and a corresponding bounce solution. In single-field tunneling, this bounce solution provides the field profile. In the multi-field case, it gives the profile of each field, and the tunneling path in the field space can also be obtained from it. We also optimize the calculation of the super-cooling case by focusing on the effective potential near the false vacuum and the barrier, to avoid errors caused by the tunneling point being too small relative to the search range. We have provided a comparison between our package and others in terms of both calculation results and computation time. A comparison between the results of tunneling with and without a renormalization factor is also demonstrated.

This paper is organized as follows. In Section \ref{sectionB}, we present a detailed analysis of the NLO tunneling problems and the numerical method used to solve the NLO tunneling equation. In the next section, we provide a guide to the installation and running of our package. Then, we demonstrate how to use the package by giving concrete physical examples. Finally, conclusions and discussions, as well as the Appendix, are cast in the remaining two sections. 

\section{Shooting and path deformation method with renormalization factor}\label{sectionB}

\subsection{The necessity of tunneling Rate in NLO}
Starting with the general expression of the effective action with the derivative expansion, the detailed derivation can be found in App.\ref{appendix0}
\begin{equation}\label{eqEW}
    S_{eff}=\int d^4x \frac{Z(\phi)}{2}\partial_\mu \phi \partial^\mu\phi+V_{eff}(\phi),
\end{equation}
where $Z$ is the renormalization factor for the classical background field and $V_{eff}$ is the effective potential of the theory. 
If we rewrite them order by order then $Z=1+Z_g+Z_{g^2}+...$, and $V_{eff}=V_{tree}+V_g+V_{g^2}+...$, where the lower index $g$ represents the order of interaction coupling~\cite{Croon:2020cgk,Gould:2023ovu}. Usually, when discussing the real universe, we at least need NLO finite temperature effective potential $V_g$ to consider the temperature evolution of the universe, and, for kinetic terms, we only consider the leading order by treating $Z=1$. This lead to action
\begin{equation}
    S_{eff}=\int d^4x \frac{1}{2}\partial_\mu \phi \partial^\mu\phi+V_{eff}(\phi),
\end{equation}
as the starting point to compute the bounce action and the tunneling rate.

Nevertheless, this treatment is flawed. Consistent perturbation theory necessitates considering the NLO correction of the renormalization factor $Z_g$ when calculating the vacuum tunneling rate. This incorrect application of perturbation theory not only yields a minor NLO correction but also introduces huge uncertainty. The uncertainty arises from the gauge dependence of the vacuum tunneling rate~\cite{Metaxas:1995ab,Garny:2012cg,DiLuzio:2014bua} and brings an order of magnitude difference of the observable~\cite{Chiang:2017zbz}. So, one must treat the perturbation theory consistently. At the leading order, we can only compute the vacuum tunneling rate for 
\begin{equation}
    S_{eff}=\int d^4x \frac{1}{2}\partial_\mu \phi \partial^\mu\phi+V_{tree}(\phi),
\end{equation}
and for NLO we have
\begin{equation}
    S_{eff}=\int d^4x \frac{1+Z_g}{2}\partial_\mu \phi \partial^\mu\phi+V_{tree}(\phi)+V_{g}(\phi).
\end{equation}
Although there are some methods to avoid solving the NLO tunneling problems and obtain the gauge-independent tunneling rate by reorganizing the perturbation order through the Nielsen identity~\cite{Lofgren:2021ogg,Hirvonen:2021zej,Arunasalam:2021zrs}, designing a package for the full solution of the NLO tunneling problem, which may be suitable to check the gauge invariant of it, is still quite interesting. This paper mainly focuses on the design of the calculation package and leaves the analysis of the real NLO EWPT vacuum decay rate in our next work.

It is also worth mentioning that some phase transitions in the QCD sector would have vanished kinetic terms at the leading order~\cite{Helmboldt:2019pan,Reichert:2021cvs}. So, for those theories, a package to solve the NLO tunneling problem is also necessary. We present a concrete example of this physical situation in the last section of this paper.

\subsection{Single-field}\label{sectionB2}
\begin{figure}[h]
    \label{1dshooting}
	\centering
    \begin{tikzpicture}
	    \draw [->] (-1,0)--(8,0) node[below right] { $\sigma$ };
	    \draw [->] (0,-1.7)--(0,2.5) node[above left] { $-V_{eff}$ };
		\draw[blue,line width=1.5pt,domain=0:7,samples=100] plot(\x,{ -\x^4 / 37.5 +\x^3 /3 -\x^2});
		\draw[fill=red] (0,0) circle (2.5pt) node[above left]{$\sigma_F$};
		\draw[fill=red] (2.89,-2.17) circle (2.5pt) ;
  	\draw[fill=red] (2.89,-2.77) circle (2.5pt) ;
        \draw (2.89,0.5) node{$\sigma_b$};
        \draw[dash dot,gray,line width=1.5pt] (2.89,0)--(2.89,-2.77);
		\draw[fill=red] (6.48,1.69) circle (2.5pt) ;
        \draw[fill=red] (6.48,3.1) circle (2.5pt) ;
        \draw (6.48,-0.5) node{$\sigma_T$};
        \draw[dash dot,gray,line width=1.5pt] (6.48,0)--(6.48,3.1);
		\draw[fill=red] (5,0) circle (2.5pt) node[below right]{$\sigma_0$};
		\draw[dashed,brown,line width=1.5pt,domain=0:7,samples=100] plot(\x,{ (-4.*\x^2 - 0.8*\x^2.5 + 1.3333*\x^3 + 0.286*\x^3.5 - 0.107*\x^4 - 0.0237*\x^4.5)/4});
		\draw (7,3.3) node{$V^{Z}_{eff}$};
        \draw[fill=red] (5.5,0.81) circle (2.5pt) node[below right]{$\sigma_i$};
        \draw [->,line width=1pt] (5.35,0.81)--(5.1,0.4);
	\end{tikzpicture} 
	\caption{The blue solid line is the effective potential, in which shooting without renomalization occurs. The brown dashed line is a new potential transformed by $Z_\sigma$. The two lines have the same positions of extrema.}
\end{figure}
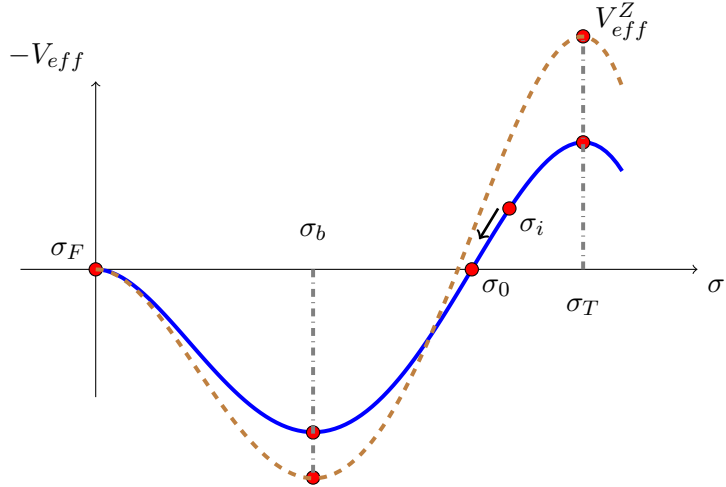

For the convenience of the numerical computations, we would like to redefine the renormalization factor $Z$ by $1/Z$. The D-dimensional Euclidean action with $\mathcal{O}(D)$ symmetry reads
\begin{equation}\label{action}
	S=4\pi \int_{0}^{\infty} \mathrm{d}r\, r^{D-1} \left[\frac{Z_{\sigma}^{-1}}{2} \left(\frac{\mathrm{d}\sigma}{\mathrm{d}r}\right)^2 +V_{eff}(\sigma)\right],
\end{equation}
where $D=4(3)$ for quantum(thermal) tunneling. The equation of motion of the action is given by
\begin{equation}\label{eq1d}
	\frac{\mathrm{d}^2 \sigma}{\mathrm{d}r^2}+\frac{D-1}{r} \frac{\mathrm{d}\sigma}{\mathrm{d}r}-\frac{1}{2}\frac{\partial \mathrm{log} Z_{\sigma}}{\partial \sigma} \left(\frac{\mathrm{d}\sigma}{\mathrm{d}r}\right)^2 = Z_{\sigma}\frac{\partial V_{eff}}{\partial \sigma},
\end{equation}
with boundary conditions
\begin{equation}
	\left. \frac{\mathrm{d}\sigma}{\mathrm{d}r}\right|_{r=0}=0, \lim_{r\to\infty}\sigma (r)=\sigma_F.
\end{equation}
Here $\sigma_F$ is the false vacuum, and the true vacuum is denoted as $\sigma_T$ below. When $Z_{\sigma}=1$, the equation of motion(\ref{eq1d}) can be reduced to 
\begin{equation}\label{eqz1}
	\frac{\mathrm{d}^2 \sigma}{\mathrm{d}r^2}+\frac{D-1}{r} \frac{\mathrm{d}\sigma}{\mathrm{d}r} = \frac{\partial V_{eff}}{\partial \sigma},
\end{equation}
and the bubble profile can be solved by the shooting method, which was first proposed by Coleman~\cite{Coleman:1977py}. Shooting method can be understood as a particle released from rest and moving in a potential equals $-V_{eff}$ with a friction term $\frac{D-1}{r} \frac{\mathrm{d}\sigma}{\mathrm{d}r}$, where $\sigma$ plays the role of position and $r$ corresponds to time, as shown in Fig.\ref{1dshooting}. 

The field value and its derivative at the ending point are fixed, as is the derivative at the initial point, but the field value at the initial point is not. So, what we have to do is to find the initial value of the field($\sigma(0)=\sigma_i$), which satisfies the ending conditions. First, we are sure that the initial point is located between the true vacuum and the point where the potential equals that of the false vacuum, which is called $\sigma_0$ here, because the particle must have more energy than the false vacuum to overcome friction. Then we can choose an arbitrary initial point between $\sigma_0$ and the true vacuum and integrate the equation of motion to the false vacuum.

The correct ending condition is that $\sigma=\sigma_F$ when $r\to\infty$. But as an ending condition of numerical integration, we choose that $\frac{\mathrm{d}\sigma}{\mathrm{d}r}\to 0$ when $\sigma = \sigma_F$, because we do not really use infinity in numerical evaluation. In this case, we use a coefficient called \texttt{RelativeAccuracy} (default value is 0.01) to control the accuracy of $\frac{\mathrm{d}\sigma}{\mathrm{d}r}\to 0$ and $\sigma = \sigma_F$. For the field itself, we treat
\begin{equation}
    |\sigma-\sigma_F|<|\sigma_b-\sigma_F|\texttt{RelativeAccuracy},
\end{equation}
as $\sigma=\sigma_F$ and for the derivative of the field, we apply an estimation to the equation of motion,with the details presented in App.\ref{appendix1}. Therefore, we treat
\begin{equation}
    \left|\frac{\mathrm{d}\sigma}{\mathrm{d}r}\right| <  \texttt{RelativeAccuracy} \left/\sqrt{\left|V_{eff}(\sigma_T)-V_{eff}(\sigma_F)\right|/D}\right..
\end{equation}
as $\frac{\mathrm{d}\sigma}{\mathrm{d}r}\to 0$. The \texttt{RelativeAccuracy} coefficient is adjustable.

However, the initial point we chose at the beginning is generally incorrect. Therefore, we will need to adjust the position of the initial point according to the results of integration. If the initial point is too close to the true vacuum, the particle carries too much energy and does not stop at the false vacuum, denoted by $\frac{\mathrm{d}\sigma}{\mathrm{d}r}$ larger than the accuracy and in the same direction as going from the false vacuum to the true vacuum. Conversely, if the initial point is too close to $\sigma_0$, the particle carries too little energy and does not reach the false vacuum, denoted by $\frac{\mathrm{d}\sigma}{\mathrm{d}r}$ change sign between the false vacuum and barrier. 

So, we can use the bisection method to find the initial point until we meet the accuracy requirement. We implement the bisection method as follows: assign weights $W_T$ and $W_0$ to $\sigma_T$ and $\sigma_0$, respectively. Then, the position of the initial point can be expressed as
\begin{equation}
    \sigma_i = \frac{W_T \sigma_T + W_b \sigma_0}{W_T+W_0}.
\end{equation}
If $\sigma$ overshoots the false vacuum, we need to search for the initial point closer to $\sigma_0$. In this case, the weights for the next step are given by
\begin{align}
    W_T^+ = 2W_T-1, \\
    W_0^+ = 2W_0+1.
\end{align}
Conversely, if $\sigma$ does not reach the false vacuum, and we need to search for the initial point closer to $\sigma_T$, the weights for the next step are given by
\begin{align}
    W_T^+ = 2W_T+1, \\
    W_0^+ = 2W_0-1.
\end{align}
The initial point for the next step is then uniformly expressed as
\begin{equation}
    \sigma_i^+ = \frac{W_T^+ \sigma_T + W_0^+ \sigma_0}{W_T^+ + W_0^+}.
\end{equation}
After repeating this process $n$ times, the accuracy of the initial point would reach$|\sigma_0-\sigma_T|/(2^n-1)$, so we can achieve arbitrarily high precision for the initial point. Therefore, we typically set an upper limit on the number of iterations for the search. If the search reaches the maximum number of iterations without the shooting process meeting the boundary condition, the program will output a message indicating that the precision of the initial point has reached a certain level, but the equation still cannot be solved. This suggests that there may be an issue with another part of the process we are trying to solve. Or, when $\frac{\mathrm{d}\sigma}{\mathrm{d}r}$ becomes sufficiently small at $\sigma=\sigma_F$, the initial point from this attempt can be considered as the correct initial point. Finally, we record $r$, $\sigma$ and $\frac{\mathrm{d}\sigma}{\mathrm{d}r}$ at each step of numerical integration, which makes up the bubble profile and substitute them into Eq.(\ref{action}) to get the Euclidean action.

When $Z_{\sigma} \neq1$, the whole process still applies, but some steps need to be modified. We can still treat the equation of motion as a particle released from rest and moving in a potential, but the potential is no longer $-V_{eff}$. We can think of it as a new potential whose derivative is $-Z_{\sigma}(\partial V_{eff}/\partial \sigma)$, and call it $-V^{Z}_{eff}$. Since this paper only discusses the case where $Z_{\sigma}$ is greater than 0 between the true and false vacua, $V^{Z}_{eff}$ has the same true vacuum, false vacuum and barrier as $V_{eff}$, because multiplying their derivatives by $Z_{\sigma}$ does not change where the derivative is 0, but the point where the potential equals that of the false vacuum changes. However, it is not efficient to solve $V^{Z}_{eff}$ in order to find the changed $\sigma_0$, and furthermore, the integration only requires the derivative of $V^{Z}_{eff}$, not $V^{Z}_{eff}$ itself. So, we broaden the search scope of the initial point to between the true vacuum and the barrier, which eliminates the need to solve for $V^{Z}_{eff}$, covers the possible range of initial points, and does not alter the criteria for adjusting the initial point. Another modification is that a new term $-\frac{1}{2}\frac{\partial \mathrm{log} Z_{\sigma}}{\partial \sigma} \left(\frac{\mathrm{d}\sigma}{\mathrm{d}r}\right)^2$ should be added to the friction term. This term may enhance the friction, causing the initial point to be closer to the true vacuum, or reduce the friction, causing the initial point to be closer to the barrier.

Sometimes, the potential and renormalization include complicated expressions that are difficult for the program to differentiate analytically, such as numerical integration. We take 1000 points between the maximum range for finding the initial point and the false vacuum and interpolate these points to obtain an approximate but analytical expression of the potential or renormalization in this interval.

The absolute value of some effective potentials can be very large, which may lead to numerical errors. To avoid this issue, we rescale the effective potential before solving the equations. After rescaling, the form of the equation remains the same, and $r$ and $\sigma$ need to undergo a rescaling transformation. First, we perform the following transformation on the effective potential:
\begin{equation}
    V_{eff} \equiv \alpha^2 U,
\end{equation}
and let
\begin{equation}
    U(\sigma_b)=1.
\end{equation}
Then
\begin{equation}
    \alpha=\sqrt{V_{eff}(\sigma_b)}.
\end{equation}
Thus, the equation of motion becomes
\begin{equation}\label{eq1dnv}
	\frac{\mathrm{d}^2 \sigma}{\mathrm{d}r^2}+\frac{D-1}{r} \frac{\mathrm{d}\sigma}{\mathrm{d}r}-\frac{1}{2}\frac{\partial \mathrm{log} Z_{\sigma}}{\partial \sigma} \left(\frac{\mathrm{d}\sigma}{\mathrm{d}r}\right)^2 = \alpha^2 Z_{\sigma}\frac{\partial U}{\partial \sigma},
\end{equation}
and one can perform the transformation on $r$:
\begin{equation}
    R \equiv \alpha r.
\end{equation}
Finally, the equation of motion becomes
\begin{equation}\label{eq1dscaled}
	\frac{\mathrm{d}^2 \sigma}{\mathrm{d}R^2}+\frac{D-1}{R} \frac{\mathrm{d}\sigma}{\mathrm{d}R}-\frac{1}{2}\frac{\partial \mathrm{log} Z_{\sigma}}{\partial \sigma} \left(\frac{\mathrm{d}\sigma}{\mathrm{d}R}\right)^2 = Z_{\sigma}\frac{\partial U}{\partial \sigma}.
\end{equation}
This equation can be used to obtain the bounce solution $\sigma=f(R)$. Through the inverse transformation, we obtain the bouncing solution of the original equation as
\begin{equation}
    \sigma(r)=f(\alpha r).
\end{equation}
More importantly, the action can be computed directly by integrating the deformed bounce solution and its derivatives.
\begin{equation}
\begin{split}
    S &= \int_0^{\infty} \mathrm{d}r r^{D-1} \left[\frac{Z_{\sigma}^{-1}}{2}\left(\frac{\mathrm{d}\sigma}{\mathrm{d}r}\right)^2-\alpha^2 U(\sigma)\right]\\
    &=\int_0^{\infty} \mathrm{d}R R^{D-1} \alpha^{2-D} \left[\frac{Z_{\sigma}^{-1}}{2}\left(\frac{\mathrm{d}\sigma}{\mathrm{d}R}\right)^2 - U(\sigma)\right].
\end{split}
\end{equation}

However, when it comes to the thin-walled case, there is a computational challenge because the thickness of the wall is much smaller than the radius of the bubble. To accurately calculate the field profile at the wall, we need to choose a very small integration step size. This, in turn, results in a very long computation time when integrating over the interior of the bubble as well. We have provided an option to reduce the integration step size, making the thin-wall case, in principle, computable. We are actively looking for methods to optimize computation for thin-wall scenarios in the future.

\subsection{multi-field}
We adopt the path deformation method proposed in~\cite{Wainwright:2011kj} to handle multi-field problems. We will provide a description of how this method works and how we include the renormalization factor into path deformation below.

When the model contains multiple fields, both the effective potential and the renormalization factor may involve multiple fields.
\begin{align}
    V_{eff}=V(\boldsymbol{\sigma})=V(\sigma_1, \sigma_2,...)\\
    Z_{\sigma}=Z(\boldsymbol{\sigma})=Z(\sigma_1, \sigma_2,...).
\end{align}
Accordingly, the equation is
\begin{equation}
    \frac{\mathrm{d}^2 \boldsymbol{\sigma}}{\mathrm{d}r^2}+\frac{D-1}{r} \frac{\mathrm{d}\boldsymbol{\sigma}}{\mathrm{d}r}-\frac{1}{2} \boldsymbol{\nabla}_{\sigma}\mathrm{log} Z_{\sigma} \left(\frac{\mathrm{d}\boldsymbol{\sigma}}{\mathrm{d}r}\right)^2 = Z_{\sigma}\boldsymbol{\nabla}_{\sigma} V_{eff}.
\end{equation}
For tunneling to occur, there still needs to be two minima of the effective potential
\begin{equation}
    \boldsymbol{\nabla}_{\sigma} V_{eff}|_{\boldsymbol{\sigma}=\boldsymbol{\sigma}_F}=\boldsymbol{\nabla}_{\sigma} V_{eff}|_{\boldsymbol{\sigma}=\boldsymbol{\sigma}_T}=0.
\end{equation}
The lower potential minimum is referred to as the true vacuum, denoted as $\boldsymbol{\sigma}_T$; the higher potential minimum is called the false vacuum, denoted as $\boldsymbol{\sigma}_F$. We consider that tunneling occurs along a path in the field space, represented by the parametric equation $\boldsymbol{\sigma}(x)$. For convenience in later calculations, we also require the parametric equation to satisfy
\begin{equation}
    \left|\frac{\mathrm{d}\boldsymbol{\sigma}}{\mathrm{d}x}\right|=1.
\end{equation}
Therefore, we choose the path length as the parameter. For example, first, we select the initial path to be a straight line, with the parameter value at the false vacuum set to 0, the parametric equation can be written as
\begin{equation}
	\boldsymbol{\sigma}(x)=\boldsymbol{\sigma}_{F}+\frac{\boldsymbol{\sigma}_T - \boldsymbol{\sigma}_F}{\left|\boldsymbol{\sigma}_T - \boldsymbol{\sigma}_F\right|}x.
\end{equation}

Along this path, the effective potential and renormalization as functions of the parameter are denoted by 
\begin{align}
	V(x)=V_{eff}(\boldsymbol{\sigma}(x))\\
	Z(x)=Z_{\sigma}(\boldsymbol{\sigma}(x)),
\end{align}
which is equivalent to a single-field effective potential and renormalization. Once we have a path (which does not necessarily have to be the path that we ultimately solve on), we can decompose the equation into components along the path and perpendicular to the path:
\begin{align}
    \frac{\mathrm{d}^2 x}{\mathrm{d}r^2}+\frac{D-1}{r} \frac{\mathrm{d}x}{\mathrm{d}r}-\frac{1}{2}\frac{\mathrm{d\,log} Z_{\sigma}}{\mathrm{d} x} \left(\frac{\mathrm{d}x}{\mathrm{d}r}\right)^2 = Z_{\sigma}\frac{\mathrm{d} V_{eff}}{\mathrm{d} x},
    \\
    \left(\frac{\mathrm{d}x}{\mathrm{d}r}\right)^2 \frac{\mathrm{d}^2 \boldsymbol{\sigma}}{\mathrm{d}x^2} -\frac{1}{2}\frac{\mathrm{d}^2 x}{\mathrm{d} r^2}\boldsymbol{\nabla}_{\perp}\left(\mathrm{log} Z_{\sigma}\right) =Z_{\sigma}\boldsymbol{\nabla}_{\perp}V_{eff},
\end{align}
where $\boldsymbol{\nabla}_{\perp}$ represents the derivative perpendicular to the path. It can be seen that the equation along the path is the same as the single-field tunneling equation. Moreover, the effective potential along the path also has the same characteristics as discussed in the previous section, namely, it has two minima. Therefore, we can apply the method from the previous section to obtain the bounce solution $x(r)$ and the action along this path. Next, we only need to determine which path satisfies the perpendicular component equation. Here, "perpendicular force" acting on the path can be defined as
\begin{equation}\label{normf}
    \boldsymbol{N}\equiv\left(\frac{\mathrm{d}x}{\mathrm{d}r}\right)^2 \frac{\mathrm{d}^2 \boldsymbol{\sigma}}{\mathrm{d}x^2} -\frac{1}{2}\frac{\mathrm{d}^2 x}{\mathrm{d} r^2}\boldsymbol{\nabla}_{\perp}\left(\mathrm{log} Z_{\sigma}\right) -Z_{\sigma}\boldsymbol{\nabla}_{\perp}V_{eff}.
\end{equation}
If the path we find is correct, the perpendicular force must be zero. However, we cannot expect to find the correct path right from the start, so we now consider the case where the path deviates from the correct path.

We first consider the leading order(LO) case ($Z_\sigma = 1$), in which the $\frac{1}{2}\frac{\mathrm{d}^2 x}{\mathrm{d} r^2}\boldsymbol{\nabla}_{\perp}\left(\mathrm{log} Z_{\sigma}\right)$ term is zero, and the third term is equal to $-\boldsymbol{\nabla}_{\perp}V_{eff}$. The negative perpendicular gradient term is equivalent to pushing the path towards the path of the minimum of the effective potential, which aligns with our expectation of minimizing the action and maximizing the tunneling rate. The first term gives a modification to this path based on the current bounce on this path.
By selecting $n$ points $\{\boldsymbol{\sigma}_{i}\}$ along this path (the program defaults to 100 points), each term in equation (\ref{normf}) can be computed at each point: $x$ being the path length combined by $|\boldsymbol{\sigma}_{i+1} - \boldsymbol{\sigma}_{i}|$, together with $V_{eff}$ at$\{\boldsymbol{\sigma}_{i}\}$, provide $x(r)$ by calling the single-field program. $\left(\frac{\mathrm{d}x}{\mathrm{d}r}\right)^2$ and $\frac{\mathrm{d}^2 x}{\mathrm{d} r^2}$ can be directly obtained by solving for $r$ from $x(r) = x_i$ and substituting $r$ into $x'(r)$ and $x''(r)$; $\frac{\mathrm{d}^2 \boldsymbol{\sigma}}{\mathrm{d}x^2}$ can be obtained by performing a finite difference between a point and its neighboring points; $\boldsymbol{\nabla}_{\perp}V_{eff}$ can be computed through
\begin{equation}
	\boldsymbol{\nabla}_{\perp}=\hat{\boldsymbol{e}}_{\perp} \left(\hat{\boldsymbol{e}}_{\perp} \cdot \boldsymbol{\nabla}\right).
\end{equation}
The gradients can be computed as follows: If $V_{eff}$ is an analytical expression, the gradients can be directly differentiated. If it is a numerical function, the gradients must be obtained by finite differences with neighboring points. The vector $\hat{\boldsymbol{e}}_{\perp}$ represents the unit vector perpendicular to the path at that point, and it can be computed from
\begin{equation}
	\hat{\boldsymbol{e}}_{\perp} \cdot (\boldsymbol{\sigma}_{i+1} - \boldsymbol{\sigma}_{i}) = 0.
\end{equation}
where $(\boldsymbol{\sigma}_{i+1} - \boldsymbol{\sigma}_{i})$ is the vector difference between the $i$-th point and its neighboring point and is considered the tangent vector of the path at the $i$-th point.
After the perpendicular force $\boldsymbol{N}_i$ is obtained at each point $\boldsymbol{\sigma}_{i}$, we move each point a small distance along $\boldsymbol{N}_i$:
\begin{equation}
	\boldsymbol{\sigma}_{i}^+=\boldsymbol{\sigma}_{i}+\alpha \boldsymbol{N}_i.
\end{equation}
Afterward, the curve formed by $\{\boldsymbol{\sigma}_{i}^+\}$ becomes the new path. The degree of path deformation, $\alpha$, should be chosen to a suitable value, especially not too large. An excessively large $\alpha$ may cause errors in the evaluation. Here, we refer to the feedback method used in ~\cite{Wainwright:2011kj} to control the size of $\alpha$. Specifically, after the path points are moved, the difference between the perpendicular forces $\{\boldsymbol{N}_{i}^+\}$ calculated from $\{\boldsymbol{\sigma}_{i}^+\}$ and the original forces $\{\boldsymbol{N}_{i}\}$ should not exceed 10\%. If it does, $\alpha$ is reduced. As $\boldsymbol{N}\to 0$, any small movement of the path causes a large change in $\boldsymbol{N}$ relative to itself, making $\alpha$ very small, and we think the path has reached the correct position.

\begin{figure}[!htbp]
	\centering
	\begin{tikzpicture}
		\draw[rounded corners=14] (-1,0) rectangle (1,1);\draw(0,0.5) node { Start };
		\draw[-{latex}] (0,0)--(0,-1);
		\draw(-2,-1) rectangle (2,-2);\draw(0,-1.5) node { Choose initial path };
		\draw[-{latex}] (0,-2)--(0,-3);
		\draw(-2,-3) rectangle (2,-4);\draw(0,-3.5) node { Solve 1-d equation };
		\draw[-{latex}] (0,-4)--(0,-5);
		\draw(-2,-5) rectangle (2,-6.5);\draw(0,-5.5) node { Calculate $\boldsymbol{N}$ };\draw(0,-6) node { by current solution };
		\draw[-{latex}] (0,-6.5)--(0,-7.5);
		\draw(0,-7.5)--(2,-8)--(0,-8.5)--(-2,-8)--cycle;\draw(0,-8) node { $\boldsymbol{N}=0$ };
		
		\draw[-{latex}] (0,-8.5)--(0,-9.5);\draw (-0.5,-9) node { YES };
		\draw[rounded corners=14] (-1,-9.5) rectangle (1,-10.5);\draw(0,-10) node { End };
		
		\draw[-{latex}] (2,-8)--(3.5,-8);\draw (2.75,-7.75) node { NO };
		\draw(3.5,-7.25) rectangle (7,-8.75);\draw(5.25,-7.75) node { Deform the path}; \draw(5.25,-8.25) node{ along $\boldsymbol{N}$ };
		\draw[-{latex}] (7,-8)--(8,-8);
		\draw(8,-7.25) rectangle (12,-8.75);\draw(10,-7.75) node { Calculate $\boldsymbol{N}$ };\draw(10,-8.25) node { by current solution };
		\draw[-{latex}] (10,-7.25)--(10,-6);
		\draw(10,-6)--(12,-5.5)--(10,-5)--(8,-5.5)--cycle;\draw(10,-5.5) node { $\boldsymbol{N}=0$ };
		
		\draw[-{latex}] (10,-5)--(10,-3.5)--(2,-3.5);\draw (10.5,-4.5) node { YES };
		
		\draw[-{latex}] (8,-5.5)--(5.25,-5.5)--(5.25,-7.25);\draw (7,-5.75) node { NO };
	\end{tikzpicture}
	\caption{Multi-Field Algorithm Program Flowchart: The main structure consists of a large loop and a small loop. The large loop is the strict process for finding the correct path, while the small loop is an approximation process that omits the re-solving of the equation along the path to reduce computational cost.}
	\label{multifieldprogram}
\end{figure}
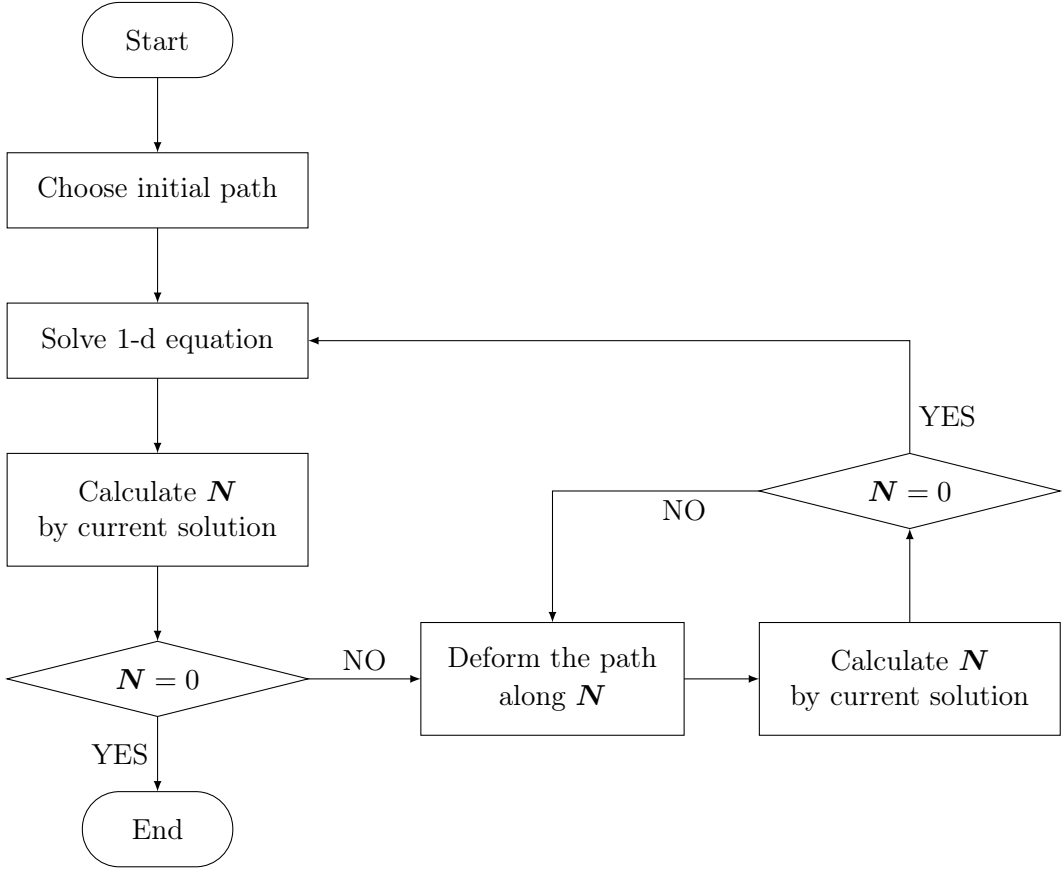

Strictly speaking, after each deformation of the path, both the effective potential and the renormalization factor along the path will change, leading to changes in the single-field bounce solution. However, in practical calculations, to reduce computational costs, we continue to use the same single-field solution to drive the path deformation until the path reaches the one where the perpendicular force is zero, or the number of deformations is sufficiently large. This is based on the assumption that small changes in the path will not introduce significant changes in the single-field solution. This approximation does not affect the final determination of whether the path is correct because after the perpendicular force becomes zero, we always verify the single-field bounce solution for the current path, as shown in \ref{multifieldprogram}. This ensures that all parameters in equation (\ref{normf}) are consistent with the current path.

Next, we discuss the NLO effects. First, the added gradient term of $\mathrm{log}Z_\sigma$ can be expressed as $\boldsymbol{\nabla}_{\perp}Z_\sigma /Z_\sigma$, meaning that the gradient of $Z_\sigma$ directly alters the perpendicular force. If the gradient of $Z_\sigma$ is large while the value of $Z_\sigma$ itself is relatively small, this term will have a significant impact on the path. Second, the gradient term of the effective potential is multiplied by $Z_\sigma$, and since $Z_\sigma$ is not constant, its distribution will exert a non-uniform influence on the perpendicular force, thereby modifying the path. Finally, due to the inclusion of $Z_\sigma$, the single-field bounce solution changes, which affects the terms $\left(\frac{\mathrm{d}x}{\mathrm{d}r}\right)^2$ and $\frac{\mathrm{d}^2 x}{\mathrm{d} r^2}$, thus further influencing the perpendicular force. Combining the changes in the path and the single-field solution along the path, $Z_\sigma$ will ultimately lead to a modification in the action.

\subsection{Super-Cooling}\label{subsubsectionsc1d}
We broaden the search scope of the initial point to between the true vacuum and the barrier in the single-field shooting, but there is also a situation in which the search scope should be narrower, which is known as a super-cooling phase transition. In this case, the barrier is very close to the false vacuum and very far from the true vacuum, and the absolute value of the potential of the true vacuum is much greater than the difference between the potential of the false vacuum and of the barrier. The initial point must be much closer to the false vacuum than the true vacuum because it does not need so much more energy to reach a point where the potential is the same, even though there is a friction term. So we give a search scope that is 10 times the difference between the false vacuum and the barrier, which is between $\sigma_b$ and $\sigma_b + 10 (\sigma_b - \sigma_F)$ if $\sigma_T$ is farther than $\sigma_b + 10 (\sigma_b - \sigma_F)$  from the barrier. In such an interval, the program can operate in the same manner as in the non-supercooling case.

In multi-field tunneling, the super-cooling case may also be encountered. Similar to single-field tunneling, only the effective potential very close to the false vacuum contributes to the tunneling. This is true not only for the shooting along the path but also because only the effective potential and the renormalization factor near the false vacuum influence the position of the path. However, the number of path points in the contributing region may be so small that it becomes difficult to reconstruct the features of the effective potential, making it extremely difficult to find the correct path and bounce solution. We place path points based on the characteristics of the first path (ranging from false vacuum to 10 times the barrier of the first path) and then deform these points to search for the correct path. This approach aims to mitigate the issue of insufficient information caused by too few path points.

\section{installation and running guide}\label{sectionA}

\subsection{Downloading and installation}
The \texttt{VacuumTunneling} package is released as a \texttt{.wl} file. It can be downloaded from \url{https://github.com/bhhua/VacuumTunneling/releases}.
\par
You can install this package by the "\textsf{Install...}" option from the "\textsf{File}" tab, choosing "\textsf{Package}" at "\textsf{Type of Item to Install}" and choosing the storage location the \texttt{.wl} file at "\textsf{Source}". Alternatively, putting this \texttt{.wl} file into
\begin{center}
	"C:\textbackslash  Users\textbackslash UserName\textbackslash AppData\textbackslash Roaming\textbackslash Mathematica\textbackslash Applications"
\end{center}
can also make it work.
\par
\subsection{Running}
This package needs to be loaded by \texttt{$<<$} command. \\
\indent\textcolor[RGB]{59,110,147}{\textsf{In[1]:=}}\;\;\texttt{$<<$"\textcolor{gray}{VacuumTunneling\`{}}"}\\
Here, we give a basic example of using this package. First, input the potential and renormalization\\
\indent\textcolor{mio}{\textsf{In[2]:=}}\;\;\textsf{\textcolor{blue}{V}[\textcolor{mvar}{x\_}] := - \textcolor{mvar}{x}\textasciicircum 2 + \textcolor{mvar}{x}\textasciicircum 3 + 2 \textcolor{mvar}{x}\textasciicircum 4;}\\
\indent\indent\indent\quad\textsf{\textcolor{blue}{Z}[\textcolor{mvar}{x\_}] := \textcolor{mvar}{x}\textasciicircum 2 + 1;}\\
and give the true vacuum and false vacuum,\\
\indent\textcolor{mio}{\textsf{In[3]:=}}\;\;\textsf{\textcolor{blue}{tv} = \textcolor{blue}{x} /. NSolve[V\textquotesingle[\textcolor{mpar}{x}] == 0, \textcolor{mpar}{x}][[1]];}\\
\indent\indent\indent\quad\textsf{\textcolor{blue}{fv} = \textcolor{blue}{x} /. NSolve[V\textquotesingle [\textcolor{mpar}{x}] == 0, \textcolor{mpar}{x}][[3]];}\\
then call the "\texttt{Tunneling}" function to evaluate the action. At least 5 arguments
\begin{center}
	\texttt{[expression of potential, expression of renormalization, field name,\\ true vacuum, false vacuum]}
\end{center} 
 should be inputted.\\
\indent\textcolor{mio}{\textsf{In[4]:=}}\;\;\textsf{\textcolor{blue}{a} = Tunneling[V[\textcolor{blue}{x}], Z[\textcolor{blue}{x}], \textcolor{blue}{x}, tv, fv]}\\
It will output an array of two elements, containing the function of field with respect to r and the action.\\
\raisebox{2\height}{\;\;\,\textcolor{mio}{\textsf{Out[4]=}}\;\;}\includegraphics[width=0.75\textwidth]{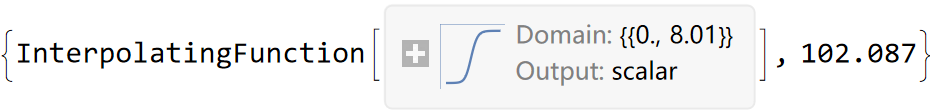}
\\
To extract the content, simply use \textsf{"a[[1]]"} or \textsf{"a[[2]]"}. For example, input
\\
\indent\textcolor{mio}{\textsf{In[5]:=}}\;\;\textsf{Plot[a[[1]][\textcolor{mpar}{r}], \{r, 0, 8.01\}]}\\
\indent\indent\indent\quad\textsf{\textcolor{blue}{Se} = a[[2]]}\\
to show the figure of the bubble profile. The plotting range can be taken as the domain of the interpolation function.
\\
\raisebox{9\height}{\;\;\,\textcolor{mio}{\textsf{Out[5]=}}\;\;}\includegraphics[width=0.55\textwidth]{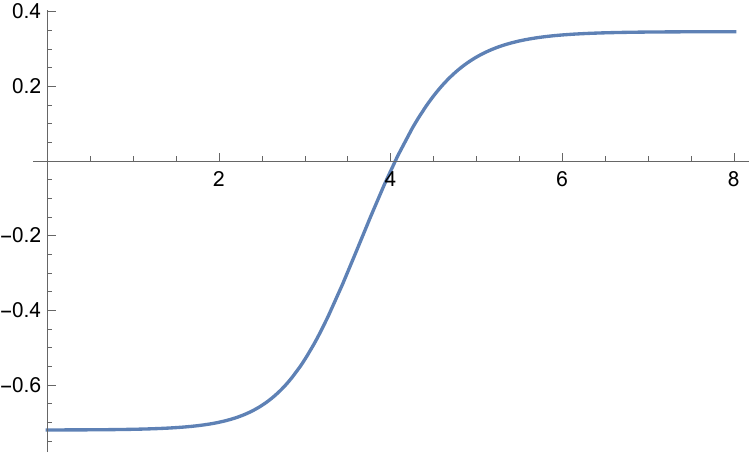}\\
\indent\indent\indent\quad\textsf{102.087}
\subsection{Options}
There are several options of the \texttt{Tunneling} function, to add these after the 5 arguments if necessary. All these options and their default values can be listed with the syntax \texttt{Options[Tunneling]}.
\par
We show the uses of the options in the following.
\begin{itemize}
\item \texttt{Dimension} corresponds to D in (\ref{action}), indicating quantum or thermal tunneling. Its default value is 4. If you are evaluating thermal tunneling, add "\texttt{Dimension->3}".
\item \texttt{TimesToFind} is the maximum times to search for the initial point of the field in single-field shooting. Its default value is 50. When the search reaches this value but still fails to meet the accuracy requirements, the function will output a message and give the bounce solution and action obtained with the initial point of the last attempt. Increasing the value of this option allows the function to further narrow the range of the initial point and may help to obtain a solution that meets the accuracy.
\item \texttt{RelativeAccuracy} defaults to 1/100, controlling the accuracy of the evaluation. Decreasing this value means the endpoint is closer to the false vacuum, and the derivative at the false vacuum is closer to 0, which requires a more accurate initial point. Therefore, reducing the value of this option to improve accuracy may significantly increase the computation time.
\item \texttt{StepScale} defaults to 1. You can use this option to manually adjust the integration step size based on the default. Increasing the value of this option for coarser integration can speed up the computation, but it will increase the error in the calculated action or even lead to failure in obtaining a solution that meets the required accuracy. Reducing the value of this option results in smaller integration steps, less integration error, and longer computation time. This may be helpful when the program cannot find a solution that meets the accuracy requirements with the default settings.
\item \texttt{NumericalPotential} and \texttt{NumericalRenormalization} both default to \texttt{False}. If the expression of potential or renormalization has a numeric part (e.g. numerical integration), you can turn its value to \texttt{True} to let the package handle it automatically. For single-field tunneling, it will take points on the potential or renormalization and interpolate them into analytical functions. For multi-field tunneling, in addition to interpolation processing, numerical differences are used to replace analytical derivatives when gradients are required.
\item \texttt{TimesToDefom} is the maximum time to do a full cycle of path deformation, which includes re-evaluating the bounce on the path. Its
default value is 20. If path deformation reaches this value but the normal force is still non-zero, the function will output a message and give the bubble profile and action of the last path.
\item \texttt{PointsNumber} is the number of points taken for path deformation when evaluating multi-field tunneling. The default value is 100. You can reduce this option to speed up the computation or increase it to improve accuracy.
\item \texttt{BarrierBetweenVacuums} When evaluating single-field tunneling, you can leave this option as its default value \texttt{Null} to let the function calculate the barrier position automatically or input the barrier to save some steps.
\end{itemize}
\section{Examples and Comparisons}
\subsection{Single-Field}
\begin{figure}
    \centering
    \includegraphics[width=0.475\linewidth]{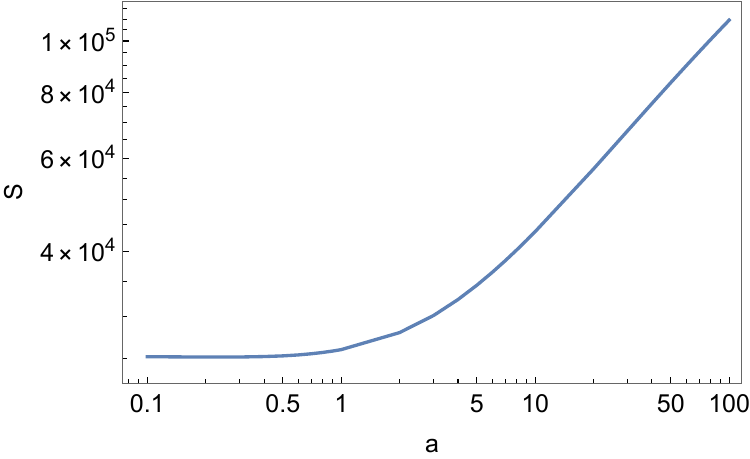}
    \includegraphics[width=0.45\linewidth]{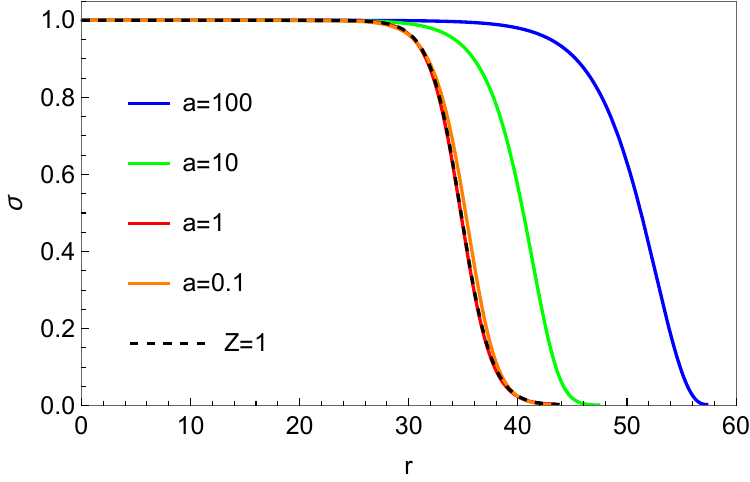}
    \caption{The left panel shows the curve of the action varying with $a$ in $Z_\sigma$. The right panel shows the bounce solutions corresponding to different values of $a$.}
    \label{fig1dz}
\end{figure}

First we demonstrate how the renormalization factor affects the bounce solution and the action. We tried this potential 
\begin{equation}\label{v1d}
	V_{eff}(\sigma)=0.25\sigma^4-0.48\sigma^3+0.22\sigma^2,
\end{equation}
whose true and false vacua is $\sigma=1$ and $\sigma=0$. We take the renormalization factor as the inverse of $Z(\phi)$ commonly used in Eq.(\ref{eqEW}), and multiply it by $a/\mathrm{Log}(a+1)$ to ensure that its average value between true and false vacua is 1. So, $Z_\sigma$ reads
\begin{equation}
    Z_\sigma=\frac{a}{\mathrm{Log}(a+1)}\frac{1}{1+a \sigma}.
\end{equation}
As long as $a>0$, $Z_\sigma$ is always positive between true and false vacua. When $a\to 0$, $Z_\sigma\to 1$. By adjusting $a$, we demonstrate the effect of different $Z_\sigma$ on the results.  As shown in Fig.\ref{fig1dz}, $Z_\sigma$ not only affects the bounce solution but also influences the action, even though the average value of $Z_\sigma$ is 1. The code we used is presented as follows:\\
\indent\textcolor[RGB]{59,110,147}{\textsf{In[1]:=}}\;\;\textsf{\textcolor{blue}{V}[\textcolor{mvar}{x\_}] := 0.25 \textcolor{mvar}{x}\textasciicircum4 - 0.48 \textcolor{mvar}{x}\textasciicircum3 + 0.22 \textcolor{mvar}{x}\textasciicircum2;}\\
\indent\textcolor[RGB]{59,110,147}{\textsf{In[2]:=}}\;\;\textsf{\textcolor{blue}{Z}[\textcolor{mvar}{x\_}, \textcolor{mvar}{a\_}] := \textcolor{mvar}{a}/Log[\textcolor{mvar}{a} + 1]/(1 + \textcolor{mvar}{a} \textcolor{mvar}{x});}\\
\indent\textcolor[RGB]{59,110,147}{\textsf{In[3]:=}}\;\;\textsf{\textcolor{blue}{b} = Tunneling[V[\textcolor{blue}{x}], Z[\textcolor{blue}{x}, 1], \textcolor{blue}{x}, 1, 0]}

This package can also be used to solve tunneling equations without renormalization factors. We compare the results obtained from this package for the case without renormalization factors with the results from existing packages to verify whether our package produces consistent results with others. As mentioned earlier, the tunneling equation without renormalization factors can be viewed as a special case where the renormalization factor $Z=1$. Therefore, to solve this case, simply input 1 in the place where $Z$ is supposed to be inputted:\\
\indent\textcolor[RGB]{59,110,147}{\textsf{In[4]:=}}\;\;\textsf{\textcolor{blue}{b1} = Tunneling[V[\textcolor{blue}{x}], 1, \textcolor{blue}{x}, 1, 0]}\\

\begin{figure}[!htbp]
	\centering
	\begin{minipage}[c]{0.49\textwidth}
		\includegraphics[width=\linewidth]{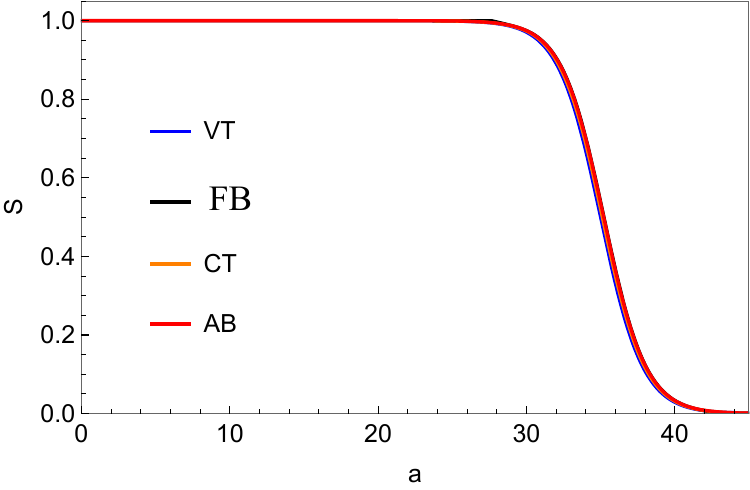}
	\end{minipage}
	\begin{minipage}[c]{0.49\textwidth}
        \begin{tabular}{|c|c|}
            \hline
            packge & action\\
            \hline
            VacuumTunneling & 25366 \\
            \hline
            CosmoTransitions & 25370\\
            \hline
            FindBounce & 25425\\
            \hline
            AnyBubble & 25379\\
            \hline
        \end{tabular}
        \label{actionpackages}
    \end{minipage}
    \caption{The left panel is the comparison between the field profile given by the program in this paper(VT) and the solution given by the \texttt{CosmoTransitions}(CT), \texttt{FindBounce}(FB) and \texttt{AnyBubble}(AB). The blue, orange, red, and green curves correspond to the solutions provided by programs \vt, \texttt{CosmoTransitions}, \texttt{FindBounce} and \texttt{AnyBubble}, respectively. The right panel is the comparison between the actions given by the programs.
    }
	\label{fig1d}
\end{figure}

The results are shown in Fig.\ref{fig1d}. In this figure, we compared our reasult with \texttt{CosmoTransitions}~\cite{Wainwright:2011kj}, \texttt{FindBounce}~\cite{Guada:2020xnz} and \texttt{AnyBubble}~\cite{Masoumi:2016wot}. The curves of the field profile almost coincide and the actions agree within 1\%. 

\subsection{Two-Fields Potential}
\begin{figure}[!htbp]
	\centering
	\begin{minipage}[c]{0.49\textwidth}
		\includegraphics[width=\textwidth]{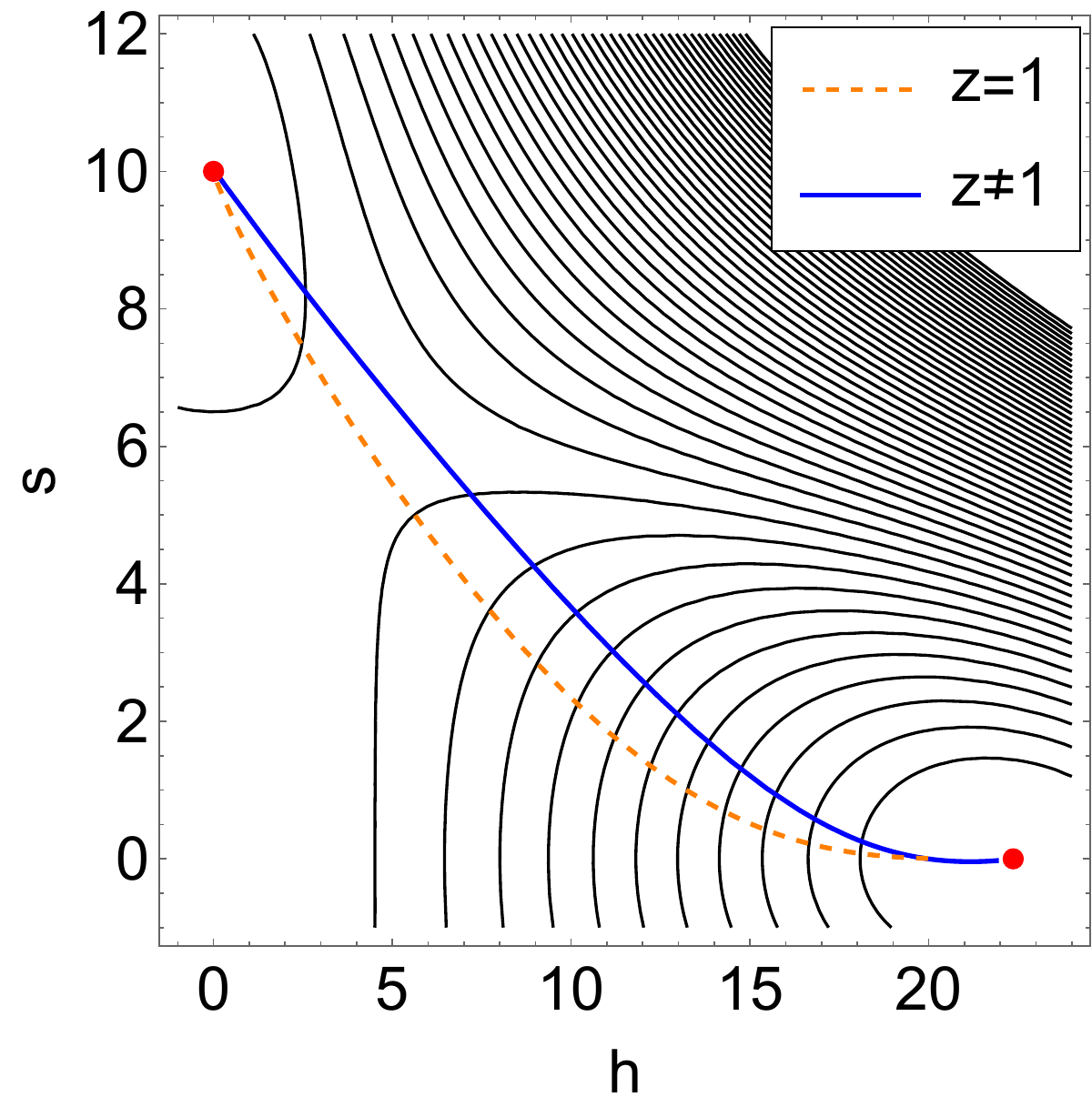}
	\end{minipage}
	\begin{minipage}[c]{0.49\textwidth}
		\includegraphics[width=\textwidth]{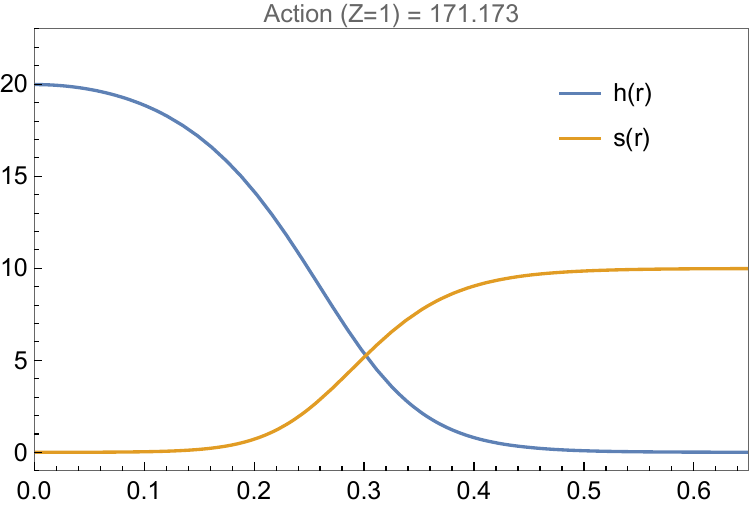}\\
		\includegraphics[width=\textwidth]{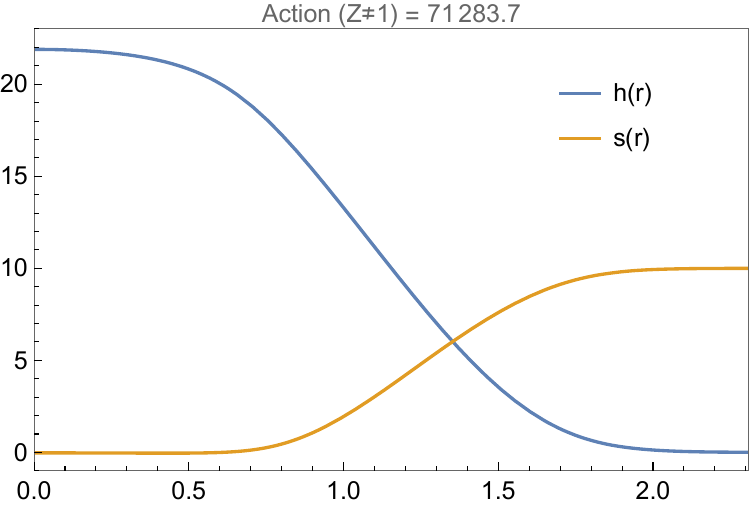}
	\end{minipage}
	\caption{The left figure is the tunneling path in the field space. The blue solid line is the path with a renormalization factor, and the orange dashed line is that without a renormalization factor. The upper right figure is the field profiles of $Z=1$. The lower right figure is the field profiles of tunneling with a renormalization factor as in Eq.(\ref{zexample2d}).}
	\label{fig2d}
\end{figure}

In addition to the single-field tunneling problem, the two-field tunneling situation is also quite common in the phase transition with the renormalization factor in the early universe. So, we also demonstrate an example and the result for tow-field tunneling problems with potential
\begin{equation}
	V_{eff}(h,s)=0.1h^4-100h^2+0.3s^4-60s^2+3h^2s^2
\end{equation}
and renormalization factor
\begin{equation}\label{zexample2d}
	Z(h,s)=\frac{0.2}{h+1}+\frac{0.1}{s+1}-\frac{0.15}{(h+1)(s+1)}.
\end{equation}
This kind of potential is commonly seen in the BSM models, and the renormalization factor is set to avoid poles in the problems. As a demonstration, we can use a similar syntax to evaluate a two-field tunneling. Input the potential and renormalization factor:\\
\indent\textcolor{mio}{\textsf{In[1]:=}}\;\;\textsf{\textcolor{blue}{V2}[\textcolor{mvar}{h\_}, \textcolor{mvar}{s\_}] := 0.1 \textcolor{mvar}{h}\textasciicircum 4 -100 \textcolor{mvar}{h}\textasciicircum 2 + 0.3 \textcolor{mvar}{s}\textasciicircum 4 -60 \textcolor{mvar}{s}\textasciicircum 2 +3 \textcolor{mvar}{h}\textasciicircum 2 \textcolor{mvar}{s}\textasciicircum 2;}\\
\indent\indent\indent\quad\textsf{\textcolor{blue}{Z2}[\textcolor{mvar}{h\_}, \textcolor{mvar}{s\_}] := 0.2 / (\textcolor{mvar}{h} + 1) + 0.1 / (\textcolor{mvar}{s} + 1) + 0.15 / ((\textcolor{mvar}{h} + 1) (\textcolor{mvar}{s} + 1));}\\
To see the general position of the vacuums, you can use the following code:\\
\indent\textcolor{mio}{\textsf{In[2]:=}}\;\;\textsf{ContourPlot[V2[\textcolor{mvar}{x}, \textcolor{mvar}{y}], \{\textcolor{mvar}{x}, -1, 24\}, \{\textcolor{mvar}{y}, -1, 12\}, Contours -> 50]}\\
as shown in Figure \ref{fig2d}. Give the true and false vacuum by
\\
\indent\textcolor{mio}{\textsf{In[3]:=}}\;\;\textsf{\textcolor{blue}{fv2} = \{\textcolor{blue}{x}, \textcolor{blue}{y}\} /. Last[FindMinimum[V2[\textcolor{blue}{x}, \textcolor{blue}{y}], \{\{\textcolor{blue}{x}, 0\}, \{\textcolor{blue}{y}, 10\}\}]]}\\
\indent\indent\indent\quad\textsf{\textcolor{blue}{tv2} = \{\textcolor{blue}{x}, \textcolor{blue}{y}\} /. Last[FindMinimum[V2[\textcolor{blue}{x}, \textcolor{blue}{y}], \{\{\textcolor{blue}{x}, 22.4\}, \{\textcolor{blue}{y}, 0\}\}]]}\\
What is different from single-field tunneling is that the field name, true vacuum, and false vacuum should be arrays with the same number of elements(\textsf{Length} in \mth).
\\
\indent\textcolor[RGB]{59,110,147}{\textsf{In[4]:=}}\;\;\textsf{\textcolor{blue}{b2} = Tunneling[V2[\textcolor{blue}{x},\textcolor{blue}{y}], Z2[\textcolor{blue}{x},\textcolor{blue}{y}], \{\textcolor{blue}{x},\textcolor{blue}{y}\}, tv2, fv2]}\\
It can also give the bubble profile and action.
\\
\raisebox{2\height}{\;\;\,\textcolor{mio}{\textsf{Out[4]=}}\;\;}\includegraphics[width=0.75\textwidth]{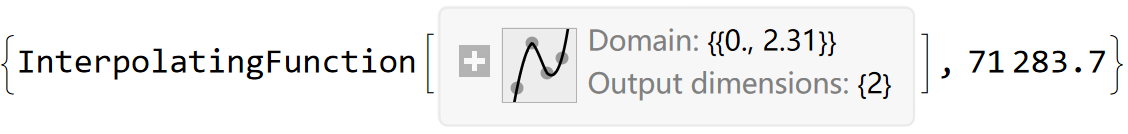}
\\
The path can be obtained by "\textsf{ParametricPlot}" function,
\\
\indent\textcolor[RGB]{59,110,147}{\textsf{In[5]:=}}\;\;\textsf{ParametricPlot[b2[[1]][\textcolor{mvar}{x}], \{\textcolor{mvar}{x}, 0, 2.31\}]}\\
If you want to see each field with respect to $r$, simply use \textsf{"Plot"} function.
\\
\indent\textcolor[RGB]{59,110,147}{\textsf{In[6]:=}}\;\;\textsf{Plot[\{b2[[1]][\textcolor{mvar}{x}][[1]], b2[[1]][\textcolor{mvar}{x}][[2]]\}, \{\textcolor{mvar}{x}, 0, 2.31\}]}\\
To evaluate tunneling without a renormalization factor, you can still input 1 where $Z$ should be.
\\
\indent\textcolor[RGB]{59,110,147}{\textsf{In[7]:=}}\;\;\textsf{\textcolor{blue}{b21} = Tunneling[V2[\textcolor{blue}{x},\textcolor{blue}{y}], 1, \{\textcolor{blue}{x},\textcolor{blue}{y}\}, tv2, fv2]}\\
As shown in Fig.\ref{fig2d}, the path in the field space at $Z\neq 1$ is modified, including the change of the initial point. Together with the effects of renormalizat]on factor on single-field shooting on the path, finally the field profiles and the action are modified.

\subsection{More than Two-Fields}\label{secMF}
Our package is designed to handle tunneling with an arbitrary number of fields. We can use the potential from~\cite{Athron:2019nbd} to test our program. The potential is formed as
\begin{equation}\label{vmmf}
	V(\phi)=\left(\left[\sum_{i=1}^{n} c_i (\phi_i -1)^2\right]-c_{n+1}\right)\left(\sum_{i=1}^{n} \phi_i^2\right)
\end{equation}
where $c_i$ takes values between 0 to 1. In~\cite{Guada:2020xnz}, there is a nice code to input this potential into \mth. This potential has extrema near each $\phi_i=0$ and $\phi_i=1$. Therefore, we adopt such a renormalization 
\begin{equation}\label{zmmf}
	Z(\phi)=\frac{1}{\prod\limits_{i=1}^{n} (a_i \phi_i +1)}
\end{equation}
which is always positive when $0<\phi_i<1$ as $a_i$ take values between 0 to 1. The coefficients $c_i$ and $a_i$ are random numbers distributed between 0 and 1, provided in App.\ref{appendix2}. We compared our results and computation time with those from other programs in the case without renormalization factor, which are demonstrated in Tab.\ref{tabmltit}. We also show the action and computation time in Tab.\ref{tabmltit}. In the test cases, the results for the action differ by less than 1\% compared to those obtained from \textsf{FindBounce}. The computation speed can be described as acceptable. As the number of fields increases, the computation time increases. Moreover, the shape of the effective potential and the renormalization coefficient also play an important role in the computation time.
\begin{table}[!htbp]
	\centering
	\begin{tabular}{ccccccc}
		\toprule
		&\multicolumn{2}{c}{VT(Z$\neq$1)}&\multicolumn{2}{c}{VT(Z=1)}&\multicolumn{2}{c}{FB(Z=1)}\\
		\cmidrule(lr){2-3} \cmidrule(lr){4-5} \cmidrule(lr){6-7}
		fields & action & time(s) & action & time(s)& action & time(s) \\
		\midrule
		2  & 625.5  & 4.59 & 243.7 & 2.25 & 245.2 & 0.016\\
		3  & 818.3  & 5.48 & 202.4 & 3.08 & 203.7 & 0.047\\
		4  & 14695  & 22.6 & 1662  & 13.6 & 1659  & 0.109\\
		5  & 926.1  & 14.0 & 122.0 & 9.97 & 122.9 & 0.141\\
		6  & 3585   & 23.3 & 399.8 & 18.3 & 401.4 & 0.125\\
		7  & 92366  & 32.1 & 2689  & 16.1 & 2703  & 0.109\\
		8  & 8880   & 28.4 & 430.7 & 19.5 & 432.6 & 0.078\\
		\bottomrule
	\end{tabular}%
	\caption{The results and computation times for tunneling without renormalization factors using our package(VT) and \texttt{FindBounce}(FB) for 2 to 8 fields, as well as the results and computation times for our package(VT) calculations with renormalization factors, are provided.}
	\label{tabmltit}
\end{table}%

\subsection{Two-Concrete Example}
To conclude this section, we present two concrete applications of our program: the supercooling phase transition with classical conformal(CSI) symmetry and the chiral phase transition in the Polyakov-Nambu-Jona-Lasino(PNJL) model.

\subsubsection{Super-Cooling Phase Transition}
\begin{figure}[!htbp]
	\centering
	\begin{minipage}[c]{0.48\textwidth}
		\includegraphics[width=\linewidth]{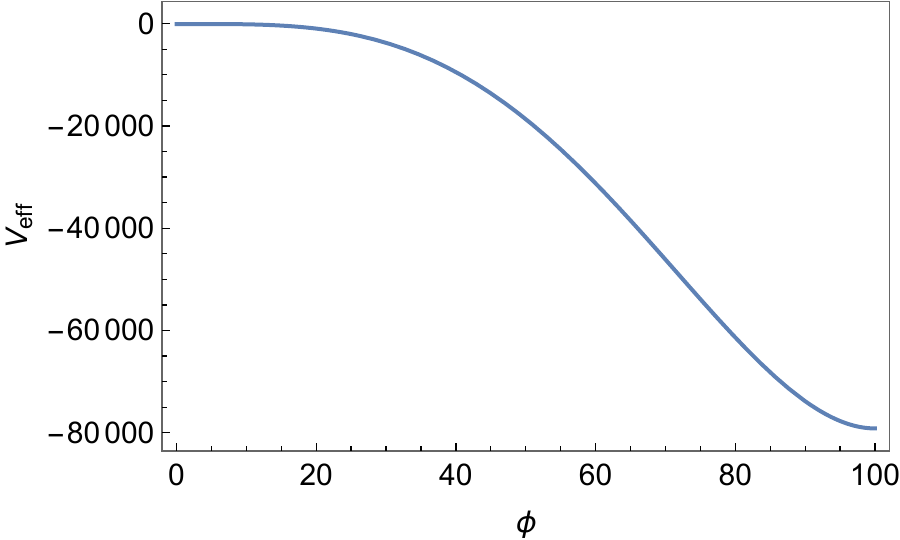}\\
		\hfill
		\includegraphics[width=0.94\linewidth]{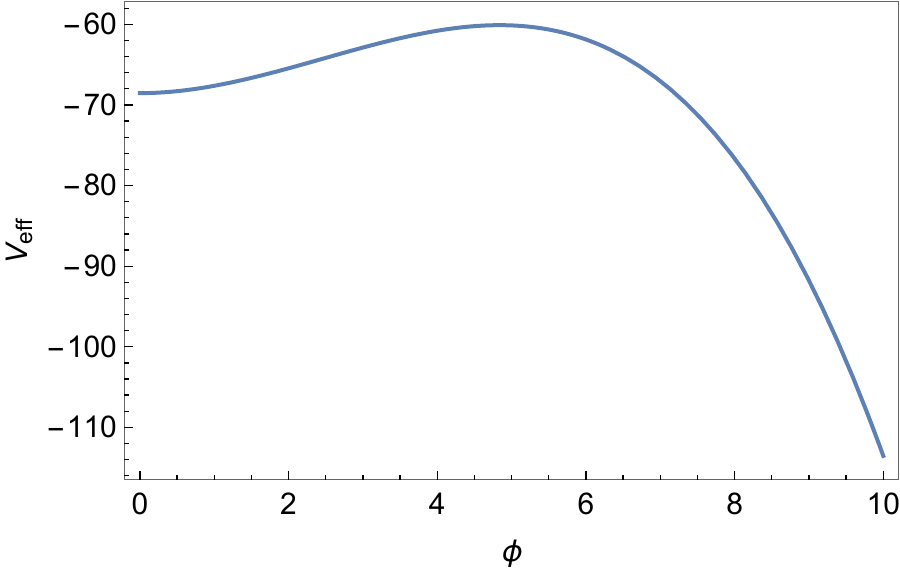}\,
	\end{minipage}
	\begin{minipage}[c]{0.45\textwidth}
		\;\includegraphics[width=0.97\linewidth]{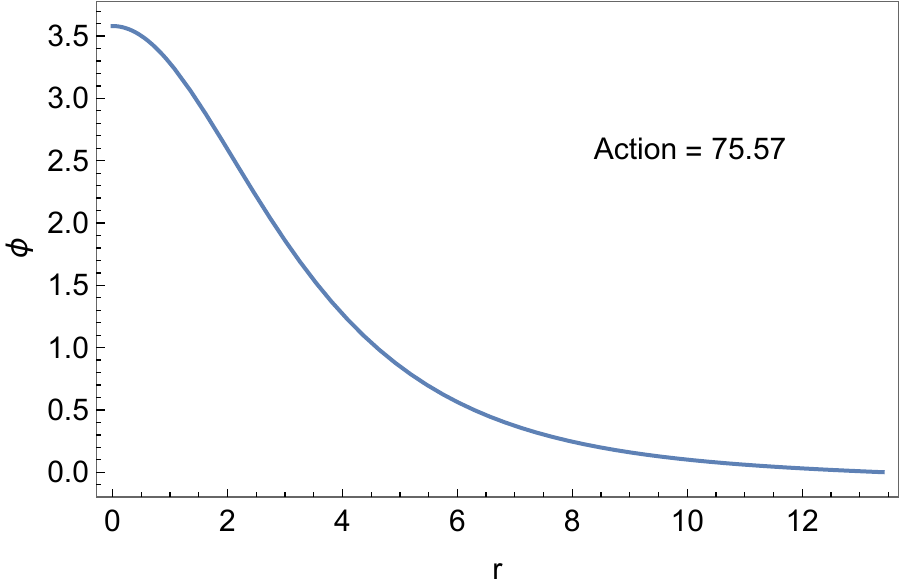}
        \includegraphics[width=\linewidth]{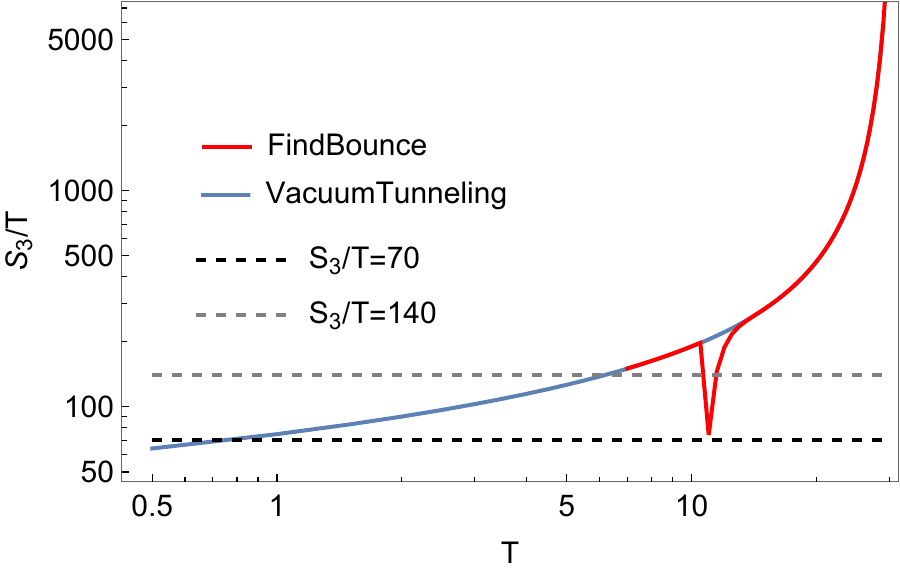}
	\end{minipage}
	\caption{Upper left and lower left: Figure of potential scaled to including the true and false vacuum and figure of potential scaled to the neighborhood of the barrier, both at $1{\rm GeV}$. Upper right: Bubble profile and action at $1{\rm GeV}$ evaluated by our package. Lower right: $S_3 /T$ with respect to $T$, evaluated by our package and \textsf{FindBounce}. Our package evaluated $S_3 /T$ for $T$ in the range from $0.5{\rm GeV}$ to $29{\rm GeV}$, while \textsf{FindBounce} evaluated $S_3 /T$ for $T$ in the range from $7{\rm GeV}$ to $29.5{\rm GeV}$ using 1000 points.}
	\label{figsc}
\end{figure}

Recently, the observation of the Nano-Grav~\cite{NANOGrav:2020bcs,NANOGrav:2023hvm,NANOGrav:2023gor} induces a lot of interesting discussion about the super-cooling phase transition in the early universe~\cite{Athron:2023mer,Wu:2023hsa,Fujikura:2023lkn,Salvio:2023ynn}. This is required since the super-cooling behavior can move the peak frequency of the GWs from the FOPT into the Nano-Hertz. In this situation, the difference between true and false vacuum is usually very large, and 
the barrier is very small compared with the whole effective potential as demonstrated 
in Fig.\ref{figsc}. We found that, for those phase transitions, the existing calculation package needs additional adjustment to calculate the correct vacuum tunneling rate. So, our package applied a modification specially designed for the super-cooling phase transition to deal with this situation.

We present a concrete, simple toy model to show this modification. The effective potential for those super-cooling phase transitions is constructed by the single scalar theory with the CSI symmetry~\cite{Kang:2020jeg, Aoki:2019mlt, Kierkla:2023von, Liu:2024fly}
\begin{equation}
    \mathcal{L}=\frac{1}{2}\partial^\mu\phi\partial_\mu\phi+\frac{1}{2}\partial^\mu X\partial_\mu X-\frac{\lambda}{4}\phi^4-\frac{g}{2}X^2\phi^2.
\end{equation}
Choosing pure scalar theory is to avoid gauge-dependent problems of the theory. $X$ field does not condense since it does not have self-interaction. In this case, the field-dependent mass of the theory is given by
\begin{equation}
    m_\phi(\phi)=3\lambda\phi^2 \quad\quad\quad m_X(\phi)=g\phi^2
\end{equation}
The effective potential of this theory reads by the Coleman-Weinberg potential in the finite temperature~\cite{Quiros:1999jp}
\begin{equation}
	\begin{split}
		V_{eff}(\lambda,g,T,\phi,\mu)&=\frac{\lambda}{4}\phi^4+\frac{1}{64\pi^2}\sum_{i=\phi}^X m_i^4\bigg[\log\frac{m_i^2(\phi)}{\mu^2}-\frac{3}{2}\bigg]+\sum_{i=\phi}^X \frac{T^4}{2\pi^2}J_B(m_i(\phi))
	\end{split}
\end{equation}
where $J_B$ is the bosonic thermal function. We do not add the daisy terms since the contribution from this term will be minor in the super-cooling phase transitions. If the coupling satisfies perturbation, one can ignore the contribution of $m_\phi$ in 1-loop effective potential since it must be 1-order smaller than the tree-level potential. In this case, to generate a non-trivial vacuum expectation value of $\phi$, one can take $\lambda = 1/(16\pi^2)$, $g=1$, $\mu = 100{\rm GeV}$ and the critical temperature is roughly $T_c\approx 31{\rm GeV}$. When temperature $T$ drops, one will expect the bubble to nucleate at nucleation temperature $T_n$. For this theory, since the existence of the CSI symmetry, this nucleation is expected to happen at very low temperatures, even at the vacuum-dominated period. For those super-cooling phase transitions, the nucleation condition for a dominated period is roughly $S_3/T_n\sim 70$~\cite{Kang:2020jeg}. To find the nucleation temperature, we present an example of \\
\indent\textcolor{mio}{\textsf{In[1]:=}}\;\;\textsf{\textcolor{blue}{Vsc}[\textcolor{mvar}{$\phi\_$}, \textcolor{mvar}{T\_}] := \textcolor{blue}{$\lambda$}/4 \textcolor{mvar}{$\phi$}\textasciicircum 4 + (\textcolor{blue}{g}\textasciicircum 2 \textcolor{mvar}{$\phi$}\textasciicircum 4)/(64 $\pi$\textasciicircum 2) (Log[(\textcolor{blue}{g} \textcolor{mvar}{$\phi$}\textasciicircum 2)/\textcolor{blue}{$\mu$}\textasciicircum 2] - 3/2) +}\\
\indent\indent\qquad\quad\textsf{\textcolor{mvar}{T}\textasciicircum 4/(2 $\pi$\textasciicircum 2) NIntegrate[\textcolor{mio}{x}\textasciicircum 2 Log[1 - Exp[-Sqrt[\textcolor{mio}{x}\textasciicircum 2 + (\textcolor{blue}{g} \textcolor{mvar}{$\phi$}\textasciicircum 2)/\textcolor{mvar}{T}\textasciicircum 2]]], \{\textcolor{mio}{x}, 0, $\infty$\}];}\\
We tried temperatures ranging from 0.5 to 29 ${\rm GeV}$ by the following code:\\
\indent\textcolor{mio}{\textsf{In[2]:=}}\;\;\textsf{\textcolor{blue}{S3oT} = Table[\{\textcolor{mio}{i}/2, Tunneling[Vsc[\textcolor{blue}{$\phi$}, \textcolor{mio}{i}/2], 1, \textcolor{blue}{$\phi$}, $\mu$, 0,}\\
\indent\indent\qquad\;\textsf{NumericalPotential -> True, Dimension -> 3][[2]]/(\textcolor{mio}{i}/2)\}, \{\textcolor{mio}{i}, 58\}];}\\
and found that the nucleation temperature is between $0.5{\rm GeV}$ and $1{\rm GeV}$. At this temperature, the difference between the true and false vacua is greater than 100 times that between the barrier and the false vacuum, and the barrier is less than $10^{-5}$ times the absolute value of the potential of the true vacuum. As mentioned in \ref{subsubsectionsc1d}, in this case, the initial point will be very close to the barrier, and searching across the entire region between the two vacua may easily result in insufficient precision to locate the initial point. As a result, errors occur in evaluation with \texttt{FindBounce}, \texttt{AnyBubble} and \texttt{CosmoTransitions}.  Our results are shown in Fig.\ref{figsc}.We also tried evaluating this potential with \texttt{FindBounce} by inputting a list of field points and found that it can work at temperatures higher than $5{\rm GeV}$ but fails at lower temperatures. Additionally, at temperatures below 7 GeV, the number of points needs to be reduced for it to work.As a more extreme example, we also tested our package with a model in which the difference between the true vacuum and vacua is $10^8$ times the difference between the false vacuum and the barrier, and it can still work.

One can also discuss the NLO tunneling effect for this scalar-induced supercooling phase transition by considering this concrete model's renormalization factor $Z_\phi$. The form of the $Z_\phi$ can be read by
\begin{equation}
    Z_\phi^{-1}=1-\frac{g^2\phi^2}{16\pi^2 M_X^2}
\end{equation}
where $M_X^2=m_X^2+\Pi_X$ and $\Pi_X=gT^2/6$ is the thermal mass. By considering this contribution, one can also compute the nucleation temperature and find that $S_3 /T$ differs less than 1\% by that without a renormalization factor since the renormalization factor is very close to 1 in the tunneling range. The situation change for the EWPT since the presence of the gauge field. In that case, the renormalization factor would change dramatically around $\phi\sim g T$, which may related to broken of low momentum expansion~\cite{Hirvonen:2021zej}. We will leave the analysis of the real NLO EWPT tunneling rate in our future works.

\subsubsection{Chiral Phase Transition}
The early universe's first-order chiral phase transition with the dark $SU(N)$ sector is very common in BSM physics. Due to the strong coupling of the $SU(N)$ sector, it is extremely difficult to derive the effective potential and discuss these phase transitions in conventional ways. One approach to studying the chiral phase transition is by using the PNJL model to obtain an effective theory for it. A major difference between chiral phase transition in the PNJL theory and ordinary phase transition is that the tunneling field lacks kinetic terms in the leading orders~\cite{Helmboldt:2019pan,Reichert:2021cvs}. So, the $Z$ factor must be considered to discuss the bounce equations, which provide a nice platform to test our programs.

We test our programs using the effective potential and the renormalization factor in~\cite{Helmboldt:2019pan}. The effective potential for the $3$-flavor $SU(3)_c$ PNJL theory is read as
\begin{equation}\label{chiralv}
\begin{split}
    V(\sigma,l,T)&=\frac{3}{8G}\sigma^2-\frac{G_D}{16G^3}\sigma^3-\frac{9}{16\pi^2}\bigg[\Lambda^4\log\bigg(1+\frac{M_\sigma^2}{\Lambda^2}\bigg)-M^4\log(1+\frac{\Lambda^2}{M_\sigma^2})\bigg]
    \\
    &-\frac{6T^4}{\pi^2}\int_0^\infty dx x^2\log\bigg(1+e^{-3\sqrt{x^2+\frac{M_\sigma^2}{T^2}}}+3le^{-\sqrt{x^2+\frac{M_\sigma^2}{T^2}}}+3l e^{-2\sqrt{x^2+\frac{M_\sigma^2}{T^2}}}\bigg)
    \\
    &+T^4\bigg[-\frac{a(T)}{2}l^2+b(T)\log(1-6l^2-3l^4+8l^3)\bigg],
\end{split}
\end{equation}
with the temperature-dependent function
\begin{equation}
    a(T)=a_0+a_1\frac{T_0}{T}+a_2\bigg(\frac{T_0}{T}\bigg)^2,\quad\quad b(T)=b\bigg(\frac{T_0}{T}\bigg)^3
\end{equation}
where $G$ and $G_D$ are the coupling parameters in the NJL model. $\sigma$ and the $l$ is the quark condensate and Polyakove loop, which act as the tunneling fields, $M_\sigma=\sigma-\frac{G_D}{8 G}\sigma^2$ is the field dependent mass for quarks, and $\Lambda$ is the cut-off for this effective theory. $l$ is the Polyakov loop, which describes the confinement effect, and $T_0$ is the confinement temperature for pure $SU(N)_c$ theory~\cite{Fukushima:2003fw}. The renormalization factor for tunneling field $\sigma$ is 
\begin{equation}\label{chiralz}
    Z_\sigma^{-1}=-9\bigg(1-\frac{G_D}{4G^2}\sigma\bigg)^2[-2A_\sigma+2B_\sigma+8C_\sigma-2 \ell_A(M_\sigma/T)+2 \ell_B(M_\sigma/T)+8 \ell_C(M_\sigma/T)]
\end{equation}
where $i$ and $l_i$ are the 1-Loop functions from the NJL effective theory. The form of the loop function and the parameter we chose to compute the vacuum decay rate are in the App.\ref{appendix3}. 

\begin{figure}[!hbtp]
    \centering
    \includegraphics[width=0.45\linewidth]{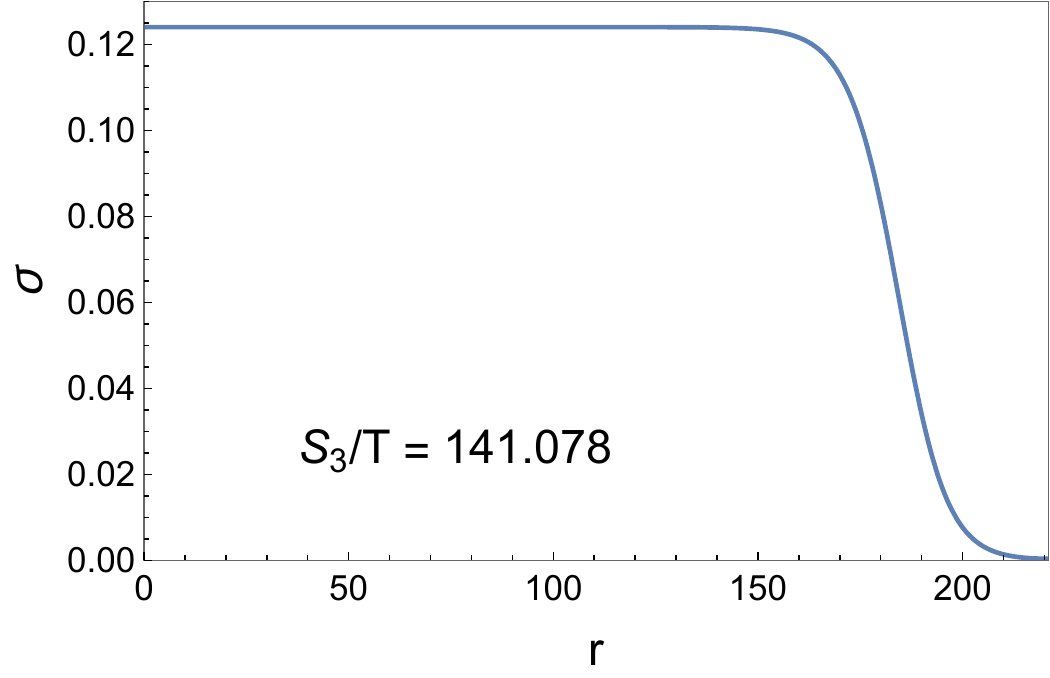}
    \includegraphics[width=0.45\linewidth]{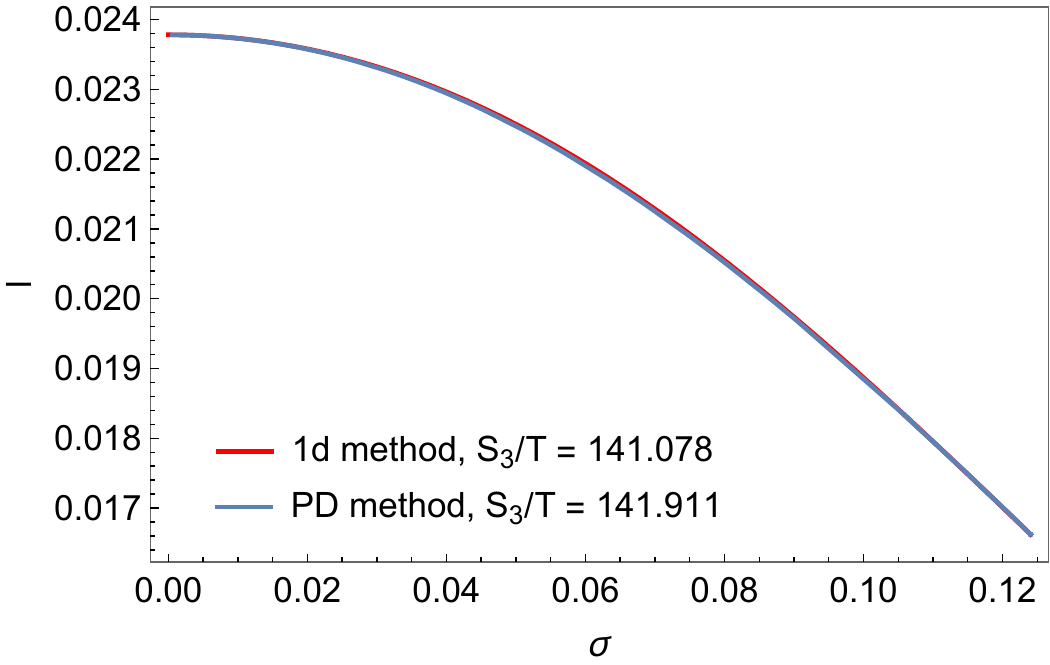}
    \caption{The left panel shows bounce solution of effective potential described by Eq.(\ref{chiralv}) and the renormalization factor described by Eq.(\ref{chiralz}), at $T=0.12182{\rm GeV}$. The $S_3 / T$ is 141.078. The right panel shows the paths found by two methods. The 1d method represents the path of the lowest potential, and the PD method represents the path found by path deformation with a renormalization factor.}
    \label{Chiral1d}
\end{figure}

Just like the approach in~\cite{Helmboldt:2019pan}, we first take the path with the lowest potential connecting the two vacua and perform single-field shooting along this path. This path can be constructed with points where $\partial V/\partial l = 0$ of each $\sigma$ at a certain temperature $T$. We take 300 $\sigma$ values between the true and false vacua here to obtain the points that form the path and perform interpolation. This path is represented as $l(\sigma,T)$, and the potential on the path is represented as $V_{eff}(\sigma, l(\sigma,T), T)$. The true and false vacua can be found by solving $\d V_{eff}(\sigma, l(\sigma,T), T)/\d \sigma = 0$, and denoted as $\sigma T$ and $\sigma F$ in the code below. Then, we can apply single-field tunneling on this path by the following code \\
\indent\textcolor{mio}{\textsf{In[1]:=}}\;\;\textsf{\textcolor{blue}{bc1} = Tunneling[Veff[\textcolor{blue}{$\sigma$}, {$l$}[\textcolor{blue}{$\sigma$}], 0.12182],  Z[\textcolor{blue}{$\sigma$}, 0.12182], \textcolor{blue}{$\sigma$}, $\sigma T$, $\sigma F$, Dimension -> 3,}\\
\indent\indent\qquad\qquad\textsf{NumericalPotential -> True,  NumericalRenormalization -> True]}\\
By using the nucleation conditions $S_3/T_n\sim 140$, We can find the nucleation temperature $T_n = 0.12182{\rm GeV}$. We choose $T_n/T_c$ as an index~\footnote{Because the parameters used in this paper may have some small differences with~\cite{Helmboldt:2019pan}, which causes the critical temperature to be not the same. We can not directly compare the nucleation temperature.} to compare the result with~\cite{Helmboldt:2019pan}. The result is very close to~\cite{Helmboldt:2019pan} with the difference smaller than 0.1 percent. 
The bounce solution and the action are shown in Fig.\ref{Chiral1d},  by\\
\indent\textcolor{mio}{\textsf{In[2]:=}}\;\;\textsf{Plot[bc1[[1]][\textcolor{mio}{x}], \{\textcolor{mio}{x}, 0, 221\}]}\\
\indent\indent\qquad\;\textsf{bc1[[2]]/0.12182}\\
However, if we take into account the modifications to the path due to the friction term and the renormalization factor, the path may differ from the curve corresponding to the lowest potential connecting the two vacua. The two-field effective potential and renormalization can be directly inputted into the \tul function to obtain the modified path, along with the corresponding action, through path deformation. The value of $l$ at the true and false vacua can be obtained by substituting $\sigma T$ and $\sigma F$ into $l(\sigma,T)$, denoted as $l T$ and $l F$ in the code below~\footnote{Because the expression of this potential and renormalization is very complicated, we took 450 × 450 points for the effective potential and performed interpolation to replace it. Additionally, to obtain convergent results, we used 500 path points.}.\\
\indent\textcolor{mio}{\textsf{In[3]:=}}\;\;\textsf{\textcolor{blue}{bc2} = Tunneling[Veff[\textcolor{blue}{$\sigma$}, \textcolor{blue}{$l$}, 0.12182],  Z[\textcolor{blue}{$\sigma$}, 0.12182], \{\textcolor{blue}{$\sigma$}, \textcolor{blue}{$l$}\}, \{$\sigma T$, $lT$\}, \{$\sigma F$, $lF$\},}\\
\indent\indent\qquad\textsf{Dimension -> 3, PointsNumber -> 500]}\\
The result is shown in Fig.\ref{Chiral1d}. The path obtained by path deformation is very close to the curve of the lowest potential connecting the two vacua, 
leading to a very small change in the action, only 0.6\%. Therefore, the nucleation temperature does not require significant modification either. This is because the Polyakov $l$ is a dimensionless field. An appropriate way to recover its mass dimension in the tunneling problem is to redefine the tunneling field of $l$ as $l\rightarrow \phi=Tl$. In this case, the field jumping value of $\phi$ in the phase transition is small compared with the $\sigma$ field. So, the tunneling problem is equivalent to the 1-dimensional tunneling problem of $\sigma$ along the lowest potential curve, which minimizes the action.

\section{Conclusion and discussion}\label{con}
Bounce action plays an important role in the FOPT of the early universe, and NLO effects provide a non-negligible modification to the bounce action. Tunneling at NLO can be described by incorporating the renormalization factor into the action. So, we present the \vt package to evaluate the bounce action with a renormalization factor. This package is based on a modified shooting and path deformation method. In single-field tunneling, by redefining the effective potential and adding an extra friction term to the equations of motion, the shooting method can still be applicable at NLO, and the action would be different from that at LO because of this modification. In the multi-field case, besides the field profile on the tunneling path being different from the LO case, the path also needs to be modified by the renormalization factor so that the final computed action can also be modified.

Our package can also calculate the LO action by setting the renormalization factor to 1. Therefore, we compared its results and computation time with those of existing packages at LO, and found that the action this package calculated agrees well with others. The computation time can be as short as a few seconds when tunneling includes $\leq$ 3 fields and tens seconds for $>$ 3 fields.

We optimize the calculation when the true vacuum is very far from the barrier relative to the false vacuum so that this package can work well in the super-cooling case. An optimization for the effective potential and renormalization as numerical functions has also been added to this package. Now, you can directly input an effective potential and renormalization with numerical functions, such as numerical integrations, to compute the action.

In this paper, we merely established a framework to compute the vacuum tunneling problems in NLO. For the thin-wall case, we have only provided a method that is theoretically computable, but it may require more computation time. We are planning to incorporate some optimizations for the thin-wall scenario into this package.
Another important task is to use this package to study how the NLO tunneling equation affected the concrete EWPT. This paper only presents a concrete example of a scalar-induced supercooling phase transition and the QCD phase transition(chiral phase transition). In both cases, the renormalization factor is non-trivial but is still a smooth and continuous function that remains positive for the domain of definition. However, for EWPT, due to the presence of the gauge field, the renormalization factor $Z$ in \eqref{eqEW} does not always be positive definite and cross zero at some field value. This behavior would cause a singularity in Eq.\eqref{eq1d} and pose challenges 
in numerical computation. So, one must regularize this singularity to do the numerical computation~\footnote{There is a concrete regularization strategy proposed by Moore and Kari~\cite{Moore:2000jw}, which is helpful to solve this problem}. We may present a concrete numerical study for the NLO-EWPT for our further work.


\noindent {\bf{Acknowledgements}}

We want to thank Prof Xiangsong Chen and Prof Zhaofeng Kang for their helpful discussions and suggestions.

\appendix
\section{The Derivation of the NLO Effective Action}\label{appendix0}
This section presents a detailed derivation of the effective action at the NLO. Starting from the definition of the effective action: the effective action is generating function of the n-point 1-PI Green's function
\begin{equation}
\begin{split}
S_{eff}(\phi_c)=\sum_{n=0}^\infty\frac{1}{n!} \int d^4x_1...d^4 x_n\bar{\phi}(x_1)...\bar{\phi}(x_n)\Gamma^n(x_1,...,x_n).    
\end{split}
\end{equation}
where $\bar{\phi}(x)$ is the classical background field and $\Gamma^n(x_1,...,x_n)$ is the n-point 1-PI Green's function. Then, one can do the Fourier transformation on the classical background field and the Green's function by
\begin{equation}
\begin{split}
    \Gamma^n(x_i)=\int\prod_{i=1}^n&\frac{d^4p_i}{(2\pi)^4}e^{ip_ix_i}(2\pi)^4\delta^4(\sum_i p_i)\Gamma^n(p_i),
    \\
    &\bar{\phi}(x)=\int\frac{d^4p}{(2\pi)^4}\wt\phi(p)e^{ipx}.
\end{split}
\end{equation}
where $\Gamma(p)$ is the 1-PI n-point function in momentum space, substituting those formulas into the expression of the effective action, one would find
\begin{equation}
\begin{split}
    S_{eff}(\phi_c)=\sum_{n=0}^\infty\frac{1}{n!}\int\frac{d^4 p_1}{(2\pi)^4}\wt\phi(-p_1)...\int\frac{d^4 p_n}{(2\pi)^4}\wt\phi(-p_n)(2\pi)^4\delta^4(p_1+...+p_n)\Gamma^n(p_i).
\end{split}
\end{equation}
Next, to give a concrete form of the effective action, we use the derivative expansion/zero momentum expansion, one has
\begin{equation}
\begin{split}
    S_{eff}(\phi_c)&=\sum_{n=0}^\infty\frac{1}{n!}\int\frac{d^4 p_1}{(2\pi)^4}\wt\phi(-p_1)|_{p_1=0}...\int\frac{d^4 p_n}{(2\pi)^4}\wt\phi(-p_n)|_{p_n=0}(2\pi)^4\delta^4(0)\Gamma^n(p)_{p=0}
    \\
    &+\frac{1}{2}\int\frac{d^4 p}{(2\pi)^4}\frac{\partial\Gamma^2(p^2)}{\partial p^2}|_{p^2=0}p^2\wt\phi(-p)\wt\phi(p)+...
\end{split}
\end{equation}
The first line is the zero-order derivative/zero-momentum expansion, which indicates the translation invariant background field, and the second line is the first-order expansion. In this case, one can define $\bar{\phi}_0(x)=\phi_c$ and the renormalization factor as
\begin{equation}\label{rewq}
\begin{split}
    &\wt\phi(p)|_{p^2=0}=\int\frac{d^4x}{(2\pi)^4}\bar{\phi}(x)e^{-ipx}=(2\pi)^4\delta^4(p)\phi_c,
    \\
    &Z=-\frac{\partial\Gamma^2(p^2)}{\partial p^2}|_{p^2=0},
\end{split}
\end{equation}
and using the definition, one finds
\begin{equation}
\begin{split}
    S_{eff}(\phi_c)=\int d^4 x \sum_{n=0}^\infty  \frac{1}{n!}\phi_c^n(2\pi)^4\Gamma^n(p)+\frac{Z(\phi_c)}{2}\int\frac{d^4 p}{(2\pi)^4}-p_\mu\wt\phi(-p)p^\mu\wt\phi(p)+...
\end{split}
\end{equation}
To extract the effective potential and kinetic terms, one can define the effective potential as
\begin{equation}
    \int d^4x V_{eff}(\phi_c)=-\int d^4 x\sum_{n=0}^\infty  \frac{1}{n!}\phi_c^n(2\pi)^4\Gamma^n(p)|_{p^2=0}
\end{equation}
and using the properties of the Fourier transformation one can obtain the kinetic terms
\begin{equation}
\begin{split}
    \int\frac{d^4 p}{(2\pi)^4}\frac{Z(\phi_c)}{2}(-p_\mu)\wt\phi(-p)p^\mu\wt\phi(p)&=\int\frac{d^4 p}{(2\pi)^4}\int d^4x d^4y \frac{Z(\phi_c)}{2} \partial_\mu\bar{\phi}(x)\partial^\mu\bar{\phi}(y)e^{-ip(x-y)}
    \\
    &=\int d^4x\frac{Z(\phi_c)}{2}\partial_\mu\bar{\phi}\partial^\mu\bar{\phi}\approx\int d^4x\frac{Z(\bar{\phi})}{2}\partial_\mu\bar{\phi}\partial^\mu\bar{\phi}.
\end{split}
\end{equation}
Finally, we find the effective action form as
\begin{equation}
    S_{eff}(\bar{\phi})=\int d^4 x [-V_{eff}(\bar{\phi})+\frac{Z(\bar{\phi})}{2}\partial_\mu\bar{\phi}\partial^\mu\bar{\phi}+...],
\end{equation}
where the ellipses represent the higher derivative/momentum expansion part.

\section{Integration Accuracy Control in Shooting}\label{appendix1}

When we have an initial point $\sigma(0)=\sigma_i$, we can get the $\sigma$ function of entire $r$ by
\begin{equation}\label{tl1}
    \sigma (r+\mathrm{d}r)= \sigma (r)+\sigma '(r)\mathrm{d}r
\end{equation}
To implement this in a program, we need to divide $r$-axis, which corresponds to time, into very small segments, each of length $\mathrm{d} r$. At each step, the current values of $\sigma$ and $\sigma'$ are used to evolve to the next $\sigma$ and $\sigma'$ after a $\mathrm{d} r$ via the equations of motion. However, since the equation of motion is a second-order differential equation, the first-order derivative of $\sigma$ with respect to $r$ cannot be obtained locally from the equation. So we have to do the same thing to the first-order derivative to get the $\sigma'$ function of the entire $r$
\begin{equation}
    \sigma' (r+\mathrm{d}r)= \sigma' (r)+\sigma ''(r)\mathrm{d}r
\end{equation}
and $\sigma''$ can be obtained locally
\begin{equation}\label{2ndd}
    \sigma ''(r)=-\frac{D-1}{r} \sigma'(r) -\frac{1}{2}(\sigma'(r))^2\left.\frac{\partial \mathrm{log} Z_{\sigma}}{\partial \sigma}\right|_{\sigma=\sigma(r)}  +\left.\frac{\partial V_{eff}}{\partial \sigma}\right|_{\sigma=\sigma(r)}.
\end{equation}
Here, although $\sigma'(r+\mathrm{d}r)$ has already been computed using $\sigma''(r)$, $\sigma(r)$ is still needed to calculate $\sigma(r+\mathrm{d}r)$; otherwise, errors may arise due to disordered integration steps.

We determine $\mathrm{d} r$ by an estimation of the LO equation Eq.\eqref{eqz1}. First we estimate the thickness of wall $\Delta r = r|_{\sigma=\sigma_F}-r|_{\sigma\approx\sigma_i}$. The first term $\frac{\mathrm{d}^2 \sigma}{\mathrm{d}r^2}$ is estimated as $\frac{\Delta\sigma}{\Delta r^2}$, the second term $\frac{D-1}{r}\frac{\mathrm{d}\sigma}{\mathrm{d}r}$ is estimated as $\frac{D-1}{\Delta r}\frac{\Delta\sigma}{\Delta r}$, and the right-hand side $\frac{\partial V_{eff}}{\partial\sigma}$ is estimated as $\frac{\Delta V_{eff}}{\Delta\sigma}$. Here we estimate $\Delta V_{eff}$ as $|V_{eff}(\sigma_T)-V_{eff}(\sigma_F)|$ and $\Delta\sigma$ as $|\sigma_T-\sigma_F|$. Then, the equation tells
\begin{equation}
    \frac{|\sigma_T-\sigma_F|}{\Delta r^2}+\frac{(D-1)|\sigma_T-\sigma_F|}{\Delta r^2} = \frac{|V_{eff}(\sigma_T)-V_{eff}(\sigma_F)|}{|\sigma_T-\sigma_F|}
\end{equation}
and finally, we get 
\begin{equation}
    \Delta r = \sqrt{D\frac{|\sigma_T - \sigma_F|^2}{|V_{eff}(\sigma_T)-V_{eff}(\sigma_F)|}}
\end{equation}
We expect the number of integration steps to be $\sim$O($10^3$) to ensure accuracy, so we take $\mathrm{d}r = \Delta r/2000$, as the action obtained with this $\mathrm{d}r$ differs by less than 1\% from the action obtained with $1/100$ of this $\mathrm{d}r$ in most of our tested cases.

After we estimate $\Delta r$, we can give the accuracy of $\frac{\mathrm{d}\sigma}{\mathrm{d}r}$. 
\begin{equation}
    \frac{\Delta\sigma}{\Delta r} = \sqrt{\frac{\left|V_{eff}(\sigma_T)-V_{eff}(\sigma_F)\right|}{D}}
\end{equation}

Finally, we have a complete algorithm that allows us to solve the entire function $\sigma(r)$, provided that we are given the starting point.
Additionally, since we already have $\sigma'$ and $\sigma''$, we can use the second-order Taylor expansion to obtain the next step’s $\sigma$
\begin{equation}\label{taylor2}
    \sigma (r+\mathrm{d}r)= \sigma (r)+\sigma '(r)\mathrm{d}r+\frac{1}{2}\sigma ''(r)\mathrm{d}r^2
\end{equation}
However, since the $\mathrm{d}r^2$ term is a higher-order small quantity when $\mathrm{d}r$ is sufficiently small, and in actual calculations, the solution obtained using only the first-order Taylor expansion is almost indistinguishable from the solution obtained using the second-order expansion, we typically default to using the first-order Taylor expansion for solving.

\section{Multi field potential and renormalization coefficients}\label{appendix2}
In \ref{secMF}, we used the potential described by Eq.(\ref{vmmf}) to test our program's capability of computing multi-field tunneling. The coefficients $c_i$ in Eq.(\ref{vmmf}) are randomly generated and listed in Tab.\ref{tablemulti}. Similarly, the coefficients $a_i$ in the renormalization described by Eq.(\ref{zmmf}) are also randomly generated and recorded in Tab.\ref{tablemulti}.
\begin{table}[!htbp]
\centering
\begin{tabular}{|c | p{14cm} |}
  \hline
  $n_\phi$ & $c_i$
  \\  \hline
  2 & 0.74201, 0.75361, 0.27580
  \\
  3 & 0.74125, 0.69371, 0.69045, 0.43404
  \\
  4 & 0.36462, 0.74735, 0.42858, 0.07069, 0.16037
  \\
  5 & 0.80423, 0.79081, 0.73818, 0.49051, 0.25790, 0.81458
  \\
  6 & 0.96192, 0.22701, 0.14752, 0.43100, 0.45173, 0.20877, 0.44505
  \\
  7 & 0.29035, 0.43834, 0.50983, 0.51155, 0.29410, 0.56446, 0.73172, 0.27239
  \\
  8 & 0.55042, 0.27161, 0.39403, 0.38593, 0.90441, 0.55011, 0.88418, 0.38275, 0.70219
  \\
  \hline
    & $a_i$
  \\
  \hline
  2 & 0.44166, 0.74307
  \\
  3 & 0.63521, 0.23525, 0.92055
  \\
  4 & 0.90959, 0.59750, 0.54071, 0.63964
  \\
  5 & 0.96977, 0.68474, 0.19992, 0.54428, 0.29792
  \\
  6 & 0.10613, 0.85147, 0.17356, 0.70600, 0.52958, 0.50874
  \\
  7 & 0.66171, 0.48841, 0.49277, 0.94936, 0.76615, 0.14302, 0.55191
  \\
  8 & 0.44741, 0.73390, 0.19297, 0.48721, 0.08316, 0.90125, 0.48916, 0.35459
  \\
  \hline
\end{tabular}
\caption{Potential coefficients $c_i$ in Eq.(\ref{vmmf}) and Renomalization coefficients $a_i$ in Eq.(\ref{zmmf})}
\label{tablemulti}
\end{table}

\section{Loop functions in the NJL model}\label{appendix3}
Theoretically, the renormalization factor $Z$ of the tunneling field can be obtained by Eq.\eqref{rewq}. To get $Z$, we at least need to compute the 1-loop 2-point function from the fermion loop of the NJL model. In this appendix, we merely summarize the 1-loop function, which is used in Eq.(\ref{chiralz}) to get the tunneling rate. The detailed derivation of those 1-loop functions can be found in~\cite{Helmboldt:2019pan} 
\begin{equation}\label{appendix31}
\begin{split}
    &A_\sigma=\frac{1}{16\pi^2}\left[\log\bigg(1+\frac{\Lambda^2}{M_\sigma^2}\bigg)-\frac{\Lambda^2}{\Lambda^2+M^2_\sigma}\right],
    \\
    &B_\sigma=-\frac{1}{32\pi^2}\frac{\Lambda^4}{(\Lambda^2+M_\sigma^2)^2},\ \ 
    C_\sigma=\frac{1}{96\pi^2}\frac{3M_\sigma^2\Lambda^4+\Lambda^6}{(\Lambda^2+M_\sigma^2)^3}.
\end{split}
\end{equation}
and
\begin{equation}\label{appendix32}
\begin{split}
    \ell_A(r) =& -\frac{1}{4\pi^2}  \int_{0}^{\infty} dx \;\biggl( \frac{x^2}{\sqrt{x^2 + r^2}^3}  \frac{1}{1 + \exp\sqrt{x^2 + r^2}} 
    \\
    & \hspace{7em} + \frac{1}{2} \frac{x^2}{(\sqrt{x^2 + r^2})^2} \frac{1}{1 + \cosh\sqrt{x^2 + r^2}} \biggr),
    \\
    \ell_B(r) =& \frac{r^2}{16 \pi^2}  \int_{0}^{\infty} dx \;\biggl( \frac{3 x^2}{\sqrt{x^2 + r^2}^5}  \frac{1}{1 + \exp\sqrt{x^2 + r^2}} +\frac{3 x^2}{2(\sqrt{x^2 + r^2})^4} \frac{1}{1 + \cosh\sqrt{x^2 + r^2}} 
    \\ 
    & \hspace{7em} + \frac{x^2}{2(\sqrt{x^2 + r^2})^5} \frac{1}{1 + \cosh\sqrt{x^2 + r^2}} \biggr),
    \\
    \ell_C(r) =& - \frac{r^4}{96 \pi^2}  \int_{0}^{\infty} d x \;\biggl( \frac{15 x^2}{\sqrt{x^2 + r^2}^7}  \frac{1}{1 + \exp\sqrt{x^2 + r^2}} +\frac{15 x^2}{2(\sqrt{x^2 + r^2})^6} \frac{1}{1 + \cosh\sqrt{x^2 + r^2}}
    \\ 
    & \hspace{7em} + \frac{3 x^2}{(\sqrt{x^2 + r^2})^5} \frac{\tanh(\sqrt{r^2+x^2}/2)}{1 + \cosh\sqrt{x^2 + r^2}} + \frac{x^2}{2(\sqrt{x^2 + r^2})^4} \frac{1}{1 + \cosh\sqrt{x^2 + r^2}}
    \\ 
    & \hspace{7em} - \frac{3 x^2}{2 (\sqrt{x^2 + r^2})^4} \frac{1}{(1 + \cosh\sqrt{x^2 + r^2})^2} \biggr).
\end{split}
\end{equation}
The parameters we choose in the effective potential Eq.\eqref{chiralv} and the 1-loop functions Eq.\eqref{appendix32} are listed in Tab.\ref{table2}. The critical temperature and the calculated nucleation temperature are also attached in Tab.\ref{table2}. 

\begin{table}[!htbp]
\begin{tabular}
{ |p{1.8cm}|p{1.9cm}|p{1.5cm}|p{1.5cm}|p{1cm}|p{1cm}|p{1cm}|p{1cm}|p{1.5cm}|p{1.5cm}|} 
 \hline
 \multicolumn{10}{|c|}{Chiral Phase Transition Parameters from NJL model} \\
 \hline
$G$[GeV$ ^{-2}$] & $G_D$[GeV$^{-5}$] & $T_0$[MeV]& $\Lambda$[MeV]& $a_0$ & $a_1$ &$a_2$ & $b$ & $T_c$[MeV] & $T_n$[MeV] \\
 \hline
3.84& -90.65& 178 & 930 & 3.51 & -2.47 & 15.2 & -1.75 & 122.25 & 121.82\\
\hline
\end{tabular}
\caption{The chiral phase transition parameters in the effective potential Eq.(\ref{chiralv}) and renormalization factor Eq.(\ref{chiralz}).}  
\label{table2}
\end{table}

\vspace{-.3cm}
\bibliographystyle{unsrt}  
\bibliography{vacuumtunneling}

\end{document}